\documentclass[letterpaper,12pt]{article}   
\usepackage{setspace}
\usepackage{osajnl2} 
\usepackage[draft]{hyperref} 

\begin{document}

\title{Tomographic reconstruction for Wide Field Adaptive Optics systems: Fourier domain analysis and fundamental limitations}


\author{Benoit Neichel,$^{1,2,*}$ Thierry Fusco,$^{1,2}$ and Jean-Marc Conan$^{1,2}$}
\address{$^1$ONERA, BP72, 92322 Chatillon Cedex, France.}
\address{$^2$Groupement d'Int\'er\^et Scientifique PHASE (Partenariat Haute r\'esolution Angulaire Sol Espace) between ONERA, Observatoire de Paris, CNRS and University Denis Diderot Paris 7}
\address{$^*$Corresponding author: benoit.neichel@onera.fr}

\begin{abstract}
Several Wide Field of view Adaptive Optics (WFAO) concepts like Multi-Conjugate AO (MCAO), Multi-Object AO (MOAO) or Ground-Layer AO (GLAO) are currently studied for the next generation of Extremely Large Telescopes (ELTs). 
All these concepts will use atmospheric tomography to reconstruct the turbulent phase volume.
In this paper, we explore different reconstruction algorithms and their fundamental limitations. 
We conduct this analysis in the Fourier domain.
This approach allows us to derive simple analytical formulations for the different configurations, and brings a comprehensive view of WFAO limitations.
We then investigate model and statistical errors and their impact on the phase reconstruction. 
Finally, we show some examples of different WFAO systems and their expected performance on a 42m telescope case. 
\end{abstract}


\maketitle 

\section{Introduction}
\label{sec_intro}
For the next generation of Extremely Large Telescopes (ELTs), several Wide Field Adaptive Optics (WFAO) concepts like Ground-layer AO (GLAO), Multi-conjuguate AO (MCAO), Laser Tomography AO (LTAO) or Multi-Object AO (MOAO) have been proposed. These systems will play a crucial role as most of them could be used as first light instruments. It is then of prime importance to understand the fundamental limitations, the possible optimizations, and the expected performance of those systems. The goal of this paper is to bring a qualitative and quantitative analysis of the most general problem of phase reconstruction and correction in multi-Guide Stars (GSs) AO. We try to point out some general trends shared by all these future systems and we propose physical interpretations of the results.\\

All these WFAO concepts have in common that they need the knowledge of the 3-dimensional turbulent volume.
The method of measuring the 3-dimensional atmospheric phase perturbations was proposed by Tallon and Foy\cite{Tallon-a-90a}: light from several Guide Stars (GSs) is used to probe the instantaneous 3-dimensional phase perturbations, the turbulent volume is then reconstructed by solving an inverse problem. This technique, called atmospheric tomography, was later improved by Johnston and Welsh\cite{Johnston-a-94a}, Ellerbroek\cite{Ellerbroek-a-94a} or Fusco\cite{Fusco-a-01a} for instance. 
The tomographic reconstruction error only depends on the GSs constellation, atmospheric conditions and Wave Front Sensor (WFS) characteristics. The geometry of the system set the first two fundamental limitations of the tomography: unseen modes and unseen turbulence.
Unseen turbulence that originates from partial pupil overlap can play a non negligible role in the global residual variance (e.g. \cite{Tokovinin-a-01b, Fusco-p-00a}), however, in the frame of ELTs this term becomes negligible \cite{Neichel-p-08a}. We then choose to 
 neglect the error due to unseen turbulence in the present study. Instead, in a first part of this paper, we focus on unseen modes and their impact on the phase estimation depending on the reconstructor choice.\\
In addition to this fundamental limitation, in a second part of this paper, we investigate an additional term due to model/statistical errors.
This term comes from a mis-knowledge of system and atmospheric conditions, which limits the tomographic reconstruction.
For instance, Conan et al.\cite{Conan-p-01a} and Tokovinin et al. \cite{Tokovinin-a-01a} have studied the sensitivity of the reconstruction error to an error on the Cn2 profile; LeLouarn and Tallon\cite{LeLouarn-a-02b} have explored the influence of the reconstructed turbulent layers altitudes compared to the real turbulent volume; Fusco et al.\cite{Fusco-a-99a}, the effect of a reconstruction on a limited number of Equivalent Layers (EL). All these studies suggest that the tomography is particularly robust to model/statistical errors. We generalize these approaches for the ELTs, and try to give a comprehensive study of the impact of these model errors.\\

Once the 3-dimensional turbulent volume is reconstructed, different types of corrections can be applied. 
The correction step consists in a projection of the reconstructed tomographic phase on the Deformable Mirrors (DMs). It only depends on the number and position of the DMs with respect to the reconstructed layers, and the optimization area.
If the corrected area is not increased compared to classical AO, only one DM conjugated to the telescope pupil is required to perform the correction (MOAO, LTAO). The main advantage of those methods is that any target selected in the FoV can be corrected: the sky coverage is significantly increased. Moreover, there are no additional terms of errors due to projection.
For a larger corrected field, a trade off between the optimization area and the DMs position is applied. This additional error term was defined by Rigaut et al. \cite{Rigaut-p-00a} and Tokovinin et al. \cite{Tokovinin-a-00a} as generalized fitting.
For instance, with one DM conjugated to the telescope pupil (GLAO), a wide field can be corrected uniformly, however the performance is dramatically limited by this error term. To reduce the impact of the generalized fitting, one must use more DMs optically conjugated to the turbulent volume.  By correcting the turbulence volume above the telescope, the efficiency of the correction is significantly improved (MCAO).
In the third part of the paper, we investigate this particular error term and its impact for ELTs.
We compare the relative performance expected for the different WFAO concepts, we explore the effect of different reconstructors, and we discuss the choice of the number of DMs. \\

We choose to conduct all these analyses in the Fourier spatial frequency domain because it allows us to perform a fast investigation of the parameter space, and it brings a comprehensive view of the impact of the different terms.
The use of spatial frequency domain techniques was initially proposed by Rigaut et al. \cite{Rigaut-p-98a}, who derived the analytical expressions for the five most fundamental limits of any Natural Guide Star (NGS)-based AO system: fitting error, angular anisoplanatism, servo-lag, WFS noise and WFS aliasing. This single conjugated NGS-AO case was later extended by Jolissaint et al. \cite{Jolissaint-a-06a}, who included the correlation between anisoplanatism and servo-lag error, and generalized the formalism for 2-dimensional systems. 
In the other hand, Tokovinin et al. \cite{Tokovinin-a-00a, Tokovinin-a-01a} have proposed an approach for evaluating, respectively, the wavefront fitting error with a limited number of DMs in a MCAO system (the so-called generalized fitting), and the tomographic reconstruction error from a limited number of noisy measurements (see also \cite{Fusco-p-02a}). Following their approach, Gavel \cite{Gavel-p-04a} has generalized the optimal tomographic reconstruction for spherical waves. Although this approach requires some simplifications, it allows to treat the so called ``focus anisoplanatism" error. Finally, Ellerbroek \cite{Ellerbroek-p-04b, Ellerbroek-a-05a} has derived an integrated approach that takes into account the fundamental error sources and their correlations for the general case of multi-GSs AO and MCAO.\\
We follow the Ellerbroek's approach, with additional developments to account for model errors and a different formalism. For sake of simplicity, we choose to not include temporal behaviors, aliasing effects or closed loop considerations. In addition, we will only consider NGS. This NGS open-loop hypothesis could appear to be restrictive, however, it allows us to obtain simple analytical formulas. Considering the complexity of WFAO systems, it is instructive to isolate specific factors in order to disentangle their individual effects. All these terms can later be included in a second step, or for a real system design for instance.

In section 2 we briefly summarize the theoretical basis of Fourier simulation and we apply this method to WFAO In Sect. 3.
Sect. 4 describes the simulation cases used thorough the paper. Sect. 5 is dedicated to the pure tomographic reconstruction and the impact of unseen frequencies. Sect. 6, investigates the impact of model errors on the tomographic phase reconstruction and Sect. 7 the impact of statistical errors. Finally, Sect. 8 focuses on the phase control and projection on DMs with some examples of different WFAO systems and their expected performance for a 42m telescope. Conclusions are given in Sect. 9.

\section{Fourier AO Modeling} 
\label{DSP}
\subsection{Introduction to Fourier simulations} 
The starting point of the Fourier method is to assume that the optical system (phase propagation, WFS measurements, DMs commands) is linear and spatially shift-invariant. In that case, all the usual operators are diagonals with respect to the spatial frequencies and simply act as spatial filters in the Fourier domain. Therefore, each equation can be written frequency by frequency. For instance, the tomographic phase reconstruction is derived and evaluated one Fourier component at a time.
In addition, our purpose is to derive regularized reconstructors which make use of the knowledge of the phase and noise statistics.
Therefore, these statistics must also be described by linear shift invariant spatial filters. This is achieved by assuming gaussian and stationary statistics. In that case, the phase and noise statistics can be entirely characterized by their 2nd order moments, more precisely, by their Power Spectrum Density (PSD).
As a consequence, the residual phase in the pupil for one direction of interest is itself fully characterized by its PSD. We will make use of this residual phase PSD to carry our analysis.
This is similar to the usual characterization of the turbulent phase by Kolmogorov / Von Karman PSD, except that we derive a PSD for the residual phase. Similarly to what is usually done for the turbulent phase, we can then construct from this residual phase PSD (i) the phase structure function, the optical transfer function, the long exposure PSF, (ii) compute the residual phase variance, (iii) generate instantaneous corrected phase screens and instantaneous AO corrected PSFs. 
The derivation of the residual phase PSD for WFAO system is described in Sect. \ref{wfao2}.\\
The main advantage of the Fourier method is that the computational complexity is significantly reduced compared to spatial domain modeling. The main limitation is that aperture-edge effects and boundary conditions, which cannot be represented by shift-invariant spatial filters, are neglected. Therefore, the Fourier modeling only applies to the ideal case of infinite aperture systems. Note that this only applies for the calculation of the PSD, as for performance evaluation we will be able to account for a finite aperture (e.g. \ref{PSDtoPSF}). The main assumption is then that all the effects of incomplete beam overlap in the upper atmospheric layers are neglected. The impact of such an assumption is discussed in \cite{Neichel-p-08a}, based on a comparison with a full E2E simulation code. Briefly, they show that as long as the different beams are superimposed in the upper layer, the error due to unseen turbulence has no impact on the tomographic reconstruction. For a 42m telescope and a typical atmospheric profile, this would correspond to a maximal GS diameter constellation around 8 arcmin.\\

\subsection{From PSD to Variance and Strehl Ratio}
\label{PSDtoPSF}
In the following, we will characterize the performance in term of residual variance and Strehl Ratio (SR). 
Following Ellerbroek \cite{Ellerbroek-a-05a} and Jolissaint \cite{Jolissaint-a-06a}, we define the residual phase variance as
 the piston-filtered integral of the residual phase PSD. This reads: 
\begin{equation}
\sigma_{res,\theta}^2 = \int\limits^\infty_0 \mbox{PSD$^{res}_{\theta}$}(f) F_p(f) df
\end{equation}
with:
\begin{equation}
F_p(f)=1-\left[ \frac{2J_1(\pi D f)}{\pi D f}\right]^2
\end{equation}
where $D$ is the telescope pupil diameter. In all this paper, D is set to 42m.\\
The SR is derived from the long-exposure PSF, which is computed from the PSD as described in Jolissaint et al. \cite{Jolissaint-a-06a}

\section{Fourier WFAO modeling} 
\label{wfao2}
The goal of this section is to derive the residual PSD when dealing with WFAO systems.

\subsection{System description and notation} 
\begin{figure}[h!]
  \begin{center}
            \includegraphics[height=5.5cm]{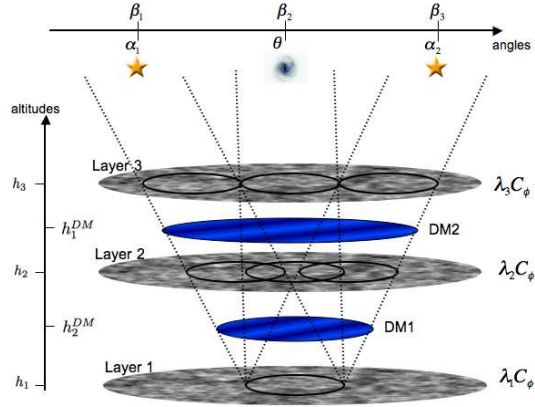}
   \end{center}
  \caption{Illustration of a system geometry. In this example, the atmosphere is simulated with three layers at altitudes $h_{n=1,2,3}$,  weighted by a $\lambda_{n=1,2,3}$ factor ($\sum\limits_{n=1}^{N_L}\lambda_n$ = 1). Two guide stars are considered in directions $\alpha_{i=1,2}$, the optimization is done in three directions $\beta_{j=1,2,3}$, the correction is perfomed with two DMs at altitudes $h_{n=1,2}^{DM}$ and the final performance is evaluated in one direction $\theta$}
    \label{description}
   \end{figure}

The atmospheric profile will be divided into $N_L$ discrete independent layers located at altitudes \{$h_n$\}.  
Each turbulent layer is described by its own power spectrum $C_{\varphi_n}$, that can be written for the $n^{th}$ layer as:
\begin{equation}
C_{\varphi_n} = \lambda_nC_{\phi}
\label{eq_cphivrai}
\end{equation}
$C_{\phi}$ is a Von-Karman power spectrum defined by:
\begin{equation}
C_{\phi}({\bf f}) = 0.023\left( \frac{1}{r_0}\right)^{5/3}\left({\bf f}^{2} + \frac{1}{L_0}\right)^{-11/6}
\end{equation}
where $r_0$ is the Fried parameter and $L_0$ is the outer scale of turbulence. In the following, we implicitly assume that all the layers have the same value of $L_0$. 
$\lambda_n$ is the fraction of turbulent energy located in the $n^{th}$ layer and is defined by: 
\begin{equation}
\lambda_n = C_n^2(h_n)\delta h_n / \sum\limits_{n=1}^{N_L}C_n^2(h_n)\delta h_n
\end{equation}
where $\delta h_n$ represents the width of each turbulent layer.\\
%
We can then write that $C_{\phi}=\sum\limits^{N_L}_{n=1} \lambda_nC_{\phi}=\sum\limits^{N_L}_{n=1} C_{\varphi_n}$. \\\\
The measurement is performed with several WFSs, each WFS looking at one Guide Star. The number of GSs/WFSs is denoted $N_{gs}$, and the associated GS positions are given by $\alpha=\{\alpha_i \}$.\\
The correction is performed by $N_{DM}$ DMs, optically conjugated to altitudes \{$h_{n}^{DM}$\}. The FoV of interest, where the correction has to be optimized, is discretized into $N_{ang}$ angles $\beta=\{\beta_j \}$. Finally, performance is computed for $N_{dir}$ direction of interests, at angles $\theta=\{\theta_k \}$. Figure \ref{description} summarizes the geometry of the system.

 \subsection{Residual phase, measurement equation and DM command}
 \label{sep}
We start with the calculation of the residual phase, expressed in the Fourier Domain. In the direction of interest $\theta$, the residual phase $\tilde{\phi}^{res}_{\theta}({\bf f})$ is given by the difference between the atmospheric phase corrugation $ \tilde{\phi}_{\theta}({\bf f})$ and the correction phase $\tilde{\phi}^{corr}_{\theta}({\bf f})$:
\begin{equation} 
\tilde{\phi}^{res}_{\theta}({\bf f}) = \tilde{\phi}_{\theta}({\bf f}) - \tilde{\phi}^{corr}_{\theta}({\bf f})
\label{eqbase}
\end{equation}
where the tilde denotes the Fourier Transform.
Assuming that propagation effects can be neglected, the turbulent phase resulting in the pupil when looking in the direction $\theta$ can be expressed as the sum of all phase perturbations. With the Fourier shift property, it leads to:
\begin{equation}
\tilde{\phi}_{\theta}({\bf f}) = \sum\limits^L_{n=1}\tilde{\varphi}_n({\bf f})e^{2j\pi {\bf f}h_n\theta}
\label{phimes}
\end{equation}
where $\tilde{\varphi}_n$ is the phase corresponding to the $n^{th}$ layer. \\\\
The correction phase is the sum of the phase perturbations introduced by the DMs. This resulting correction phase in direction $\theta$ then reads:
\begin{equation}
\tilde{\phi}^{corr}_{\theta}({\bf f}) = \sum\limits^{N_{DM}}_{n=1}\tilde{\varphi}_{n}^{DM}({\bf f})e^{2j\pi {\bf f}h_{n}^{DM}\theta}
\label{phicorr}
\end{equation}
where $\tilde{\varphi}_n^{DM}$ is the correction phase done by the $n^{th}$ DM. 
Note that each DM is assumed to correct spatial frequencies up to a cut-off frequency $f_c^{DM}$:
\begin{equation}
f_c^{DM} = \frac{N_{act}}{2 D}
\end{equation}
where $N_{act}$ is the actuator pitch as seen from the pupil plane. Beyond this cut-off frequency, no correction is applied and the corresponding phase $\tilde{\varphi}_{n}^{DM}(|{\bf f}| > f_c^{DM})$=0. \\

In, order to simplify the equations, in the following we adopt the notation $\tilde{x}$ which refers to $\tilde{x}(f)$ at a given frequency. For each frequency, Eq. \ref{phimes} and Eq. \ref{phicorr} can then be written in a vectorial form:
\begin{equation}
\begin{array}{c}
 \tilde{\phi}_{\theta} = {\bf P_{\theta}^L}{\bf \tilde{\varphi}_{turb}} \\\\
 \tilde{\phi}^{corr}_{\theta} = {\bf P_{\theta}^{DM}} {\bf \tilde{\varphi}_{DM}}
\label{eq_propa} 
\end{array}
\end{equation}
Where $\tilde{\phi}_{\theta}$ and $\tilde{\phi}^{corr}_{\theta}$ are scalar. $\bf{P_{\theta}^L}$ and $\bf{P_{\theta}^{DM}}$ are line-vectors of sizes $N_L$, respectively $N_{DM}$. ${\bf \tilde{\varphi}_{turb}}$ is a column-vector of size $N_L$ which concatenate the $N_L$ turbulent phases \{$\bf{\tilde{\varphi}_n}$\}. Similarly, $\bf{\tilde{\varphi}}_{DM}$ is a  column-vector of size $N_{DM}$ gathering the $N_{DM}$ correction phases \{$\tilde{\varphi}_n^{DM}$\}. \\\\
$\bf{P_{\theta}^L}$ projects the $N_L$ turbulent layers seen in the $\theta$ direction in the pupil plane, and $\bf{P_{\theta}^{DM}}$ projects the DM phases seen in the $\theta$ direction in the pupil. The elements of the vectors $\bf{P_{\theta}^L}$ and $\bf{P_{\theta}^{DM}}$ are simply the complex exponential appearing respectively in Eq. \ref{phimes} and Eq. \ref{phicorr}. \\\\
In the following we will assume that each measurement from the WFS is divided in two sub-measures, called hereafter measurement along $x$ and $y$. For instance, this corresponds to $x$ and $y$ slopes when considering a Shack-Hartmann (SH) sensor. In that case, the most general form for the measurement is to write it as (for a given frequency):
\begin{equation}
{\bf \tilde{\phi}^{mes}} = {\bf M P_{\alpha}^L\tilde{\varphi}_{turb}} +  {\bf b} 
\label{measure}
\end{equation}
Where ${\bf \tilde{\phi}^{mes}}$ is a (2$N_{gs}$) column-vector that concatenates the measurements coming from the $N_{gs}$ GSs. \\
${\bf P_{\alpha}^L}$ is a matrix of size $N_L$x$N_{gs}$, where we adopt a convention ``columns by rows". It projects the phases in altitudes coming from the $N_{gs}$ $\alpha_i$ directions in the pupil. Similarly to ${\bf P_{\theta}^L}$, each line of the ${\bf P_{\alpha}^L}$ matrix is filled with phase-shift complex exponential, $\theta$ being replaced by the $\alpha_i$ direction of the considered GS.\\
{\bf M} is a matrix which models the WF sensing operation (e.g. Rigaut \cite{Rigaut-p-98a}, Jolissaint \cite{Jolissaint-a-06a}, for an expression of the SH in the Fourier domain and Verinaud \cite{Verinaud-a-04a} for the pyramid one). It is a $N_{gs}$x(2$N_{gs}$) bloc diagonal matrix. Each bloc is a 2-elements column-vector modeling one WFS. The upper element is the WFS operator along $x$ and the lower element is the WFS operator along $y$. Each WFS measures frequencies up to a cut-off frequency $f_c^{WFS}$. Beyond this cut-off frequency, the WFS operator is equal to 0. The ${\bf M P_{\alpha}^L}$ matrix describes the sensitivity of the system and it will be referred as the global system interaction matrix.\\
Finally, ${\bf b}$ is a column-vector of noise associated to each GS (dimension = $2N_{gs}$). It originates from photon and detector noises. This WFS measurement noise is statistically independent for distinct GS and $x$ and $y$ components. The variance associated to this noise is assumed to be identical for all frequencies associated to a given GS, with a value denoted as $\sigma_{i}^2$ (rd$^2$) for GS number $i$. \\

Finally, one still has to express ${\bf {\tilde{\varphi}}_{DM}}$.
With linear assumptions, the most general form for the correction phases is a sum of suitably filtered measurements defined as:
\begin{equation}
{\bf {\tilde{\varphi}}_{DM}} = {\bf W\tilde{\phi}^{mes}}
\label{recons}
\end{equation}
Where {\bf W} is the phase volume reconstructor matrix of size $(2N_{gs})$ x $N_{DM}$. It converts the $(2N_{gs})$ measurements into $N_{DM}$ correction phases. Fusco et al. \cite{Fusco-a-01a} have shown that this matrix can be split in two independent matrices, written as:
\begin{equation}
{\bf W} = {\bf P_{opt}W_{tomo}}
\end{equation}
The reconstruction can thus be seen as two consecutive steps. The first one, corresponding to ${\bf W_{tomo}}$ (dimension = $2N_{gs}$ x $N_L$), provides an estimate of the turbulent phases on each reconstructed turbulent layer, it is the tomographic reconstruction. It only depends on the GSs configuration $\{\alpha_i \}$, and the atmospheric conditions. The second step, corresponding to ${\bf P_{opt}}$ (dimension = $N_L$ x $N_{DM}$), consists in a projection of the tomographic solution onto the DMs. It is a geometrical operation which provides the correction phases on the DMs from the tomographic estimated phase in the volume. It depends on the size of the scientific FoV, and on the number and positions of the DMs with respect to the reconstructed layers. In the following, we choose to investigate independently these two steps.

\subsection{Phase reconstruction: {\bf $\mbox{W}_{\mbox{tomo}}$}}
\label{wtomog}
The goal of the tomographic reconstructor is to find the best possible estimate of the phase volume from noisy GSs measurements. The phase volume estimate is written as:
\begin{equation}
{\bf \hat{\tilde{\varphi}}_{turb}} = {\bf W_{tomo} \tilde{\phi}^{mes}}
\label{eq_recons}
\end{equation}
where the hat stands for the volume estimates.
The number of sources is limited, while the number of turbulent layers is large not to say infinite, meaning that the problem is ill-conditionned and generally not directly invertible. Several methods have then been proposed to reconstruct the turbulent volume. We choose to investigate three reconstructors that are frequently used and discussed: the Minimum Mean Square Error (MMSE), the simple Least Square Estimator (LSE) and the Truncated LSE (TLSE) approach. \\

\subsubsection{MMSE reconstruction}
\label{sect_mmse}
The MMSE estimator minimizes the residual phase variance in each reconstructed layer following a quadratic criterion between actual and estimated phases:
\begin{equation}
\sigma_{res}^2=\langle || {\bf \tilde{\varphi}_{turb}} - {\bf W_{tomo}}({\bf M P_{\alpha}^L\tilde{\varphi}_{turb} + b}) ||^2 \rangle
\label{MMSEpy}
\end{equation}

In the Fourier domain, the derivation of Eq. \ref{MMSEpy} with respect to ${\bf W_{tomo}}$ is done frequency by frequency. This leads to the optimal solution defined by:
\begin{equation}
\begin{array}{l}
{\bf W^{MMSE}} = \left[ ({\bf M P_{\alpha}^{L})^T (C_b)^{-1}} {\bf M P_{\alpha}^{L}}+ {\bf C_{\varphi_n}^{-1}}\right]^{-1}
({\bf M P_{\alpha}^{L})^T(C_b)^{-1}} \\\\
\mbox{that can also be written as \cite{Tarantola-a-82a}}\\\\
{\bf W^{MMSE}} = {\bf C_{\varphi_n}}( {\bf M P_{\alpha}^{L})^T}
\left[ {\bf M P_{\alpha}^{L}C_{\varphi_n}}({\bf M P_{\alpha}^{L})^T+ C_b}\right]^{-1}
\end{array}
\label{wtomo}
\end{equation}
This reconstructor includes prior knowledge on the phase statistics and noise power spectrum by means of ${\bf C_{\varphi_n}}$ and ${\bf C_b}$.\\
As the turbulent layers are independent, for each frequency, ${\bf C_{\varphi_n}}$ is a diagonal matrix of size $N_{L}$x$N_{L}$. The diagonal components are given by the fraction of turbulent energy associated to each layer: ${\bf C_{\varphi_n}}$=diag($\lambda_1C_{\phi}$,...,$\lambda_{N_L}C_{\phi}$). \\
WFS measurement noises are statistically independent for distinct GSs and orthogonal directions, meaning that ${\bf C_b}$ is a diagonal matrix of sizes $2N_{gs}$x$2N_{gs}$. 
The noise variance is identical for all the frequencies (white noise) with a value defined by $\sigma^2_{i,(x,y)}$ (rd$^2$) for the GS number $i$. It can be different for different GSs, and for $x$ and $y$ directions: ${\bf C_b}$ $\propto$ diag($\sigma^2_{1,X}$,$\sigma^2_{1,Y}$,...,$\sigma^2_{N_{gs},X}$,$\sigma^2_{N_{gs},Y}$). \\
If we assume the same noise $\sigma^2$ for all GSs and $x$ and $y$ directions (${\bf C_b} \propto {\bf Id}$), we can factorize this noise term in the first form of ${\bf W^{MMSE}}$, and re-write Eq.\ref{wtomo} as:
\begin{equation}
{\bf W^{MMSE}} = \left[ ({\bf M P_{\alpha}^{L})^T} {\bf M P_{\alpha}^{L}}+ {\bf \sigma^2 C_{\varphi_n}^{-1}}\right]^{-1}({\bf M P_{\alpha}^{L})^T} 
\label{eq_regul}
\end{equation}
This form makes appear the regularization term ${\bf \sigma^2 C_{\varphi_n}^{-1}}$ which is nothing else that the inverse of the Signal to Noise Ratio (SNR). We will see in Sect.\ref{Sect_notionofunseen} the role of this regularization term to avoid the noise  amplification due to unseen frequencies.

\subsubsection{(T)LSE reconstruction}
The LSE estimate $\hat{\tilde{\varphi}}_{turb}$ is the one providing the best-fit to the measurements, and can be found by minimizing the following criterion :
\begin{equation}
\sigma_{res}^2 = \langle ||{\bf \tilde{\phi}_{mes}} - {\bf M P_{\alpha}^{L}}{\bf W_{tomo}}{\bf \tilde{\phi}_{mes}} ||^2 \rangle
\end{equation}
The minimization of the above criteria with respect to ${\bf W_{tomo}}$ leads to the well known solution:
\begin{equation}
{\bf W^{LSE}} = \left[ ({\bf M P_{\alpha}^{L}})^T{\bf M P_{\alpha}^{L}}\right]^{-1}({\bf M P_{\alpha}^{L})^T}
\end{equation}

It is well known and interesting to note this LSE solution corresponds to the MMSE one introduced in Eq. \ref{eq_regul} for which the regularization term have been canceled (${\bf \sigma^2 C_{\varphi_n}^{-1}}$ tends to 0).


For some frequencies, the $({\bf M P_{\alpha}^{L}})^T{\bf MP_{\alpha}^{L}}$ is not invertible or badly conditioned.
For such frequencies, Truncated Singular Value Decomposition (TSVD) is used to truncate these singular values below a pre-selected threshold. For the truncated frequencies,  ${\bf W^{LSE}}$=0 and $\hat{\tilde{\varphi}}_n$=0. The choice of the threshold is made empirically, by trying to find the best trade-off between the number of truncated frequencies and noise amplification. In simulation, it is relatively straightforward to find the optimal threshold with an exhaustive exploration of threshold levels. For each configuration, we compute the residual variance as described in \ref{PSDtoPSF} for a set of threshold levels, the optimal threshold being the one that minimizes the residual variance.
In real systems, when accounting for overheads during on-sky operations, it is not easy to scan the parameter space.
Hence all the following results for the TLSE case should be taken as optimistic ones.

\subsection{Model and statistical priors}
\label{subsect_modelstat}

Few precision on the reconstructors are important at that point.
\begin{itemize}
\item Firstly the number/altitudes of the reconstructed layers can be different from the ``real" number/altitudes of atmospheric layers as introduced in Eq. \ref{eq_cphivrai}. The choice of the number/altitudes of reconstructed layers comes from our knowledge of the atmospheric conditions. If our description of the atmospheric model is partial or wrong, we will commit an error. We will refer to this error as the ``model error".
\item Secondly the value of the statistical priors used in the MMSE reconstructor can also be different from real noise and turbulence statistics. This mis-knowledge leads to an additional error that will refer as the ``statistical error". The TLSE does not explicitly include priors on phase and noise statistics. However, the truncation level is pre-defined based on assumed models for system/atmospheric conditions. A statistical error then lead to badly tuned truncation.
\end{itemize}

We discuss the impact of the model errors in Sect. \ref{sec_modelerror}, and the statistical errors are investigated in Sect. \ref{modelerror}.

\subsection{Projection onto DMs: {\bf $\mbox{P}_{\mbox{opt}}$}}
\label{secpopt}
The tomographic reconstruction is followed by an optimal projection of the volume estimation onto the DMs. 
This can be written as:
\begin{equation}
{\bf {\tilde{\varphi}}_{DM}} = {\bf P_{opt}\hat{\tilde{\varphi}}_{turb}} 
\end{equation}
The filter that optimally projects the turbulent volume onto DMs depends on FoV optimization directions $\beta=\{\beta_i\}$ and is defined by \cite{Ellerbroek-a-94a, Fusco-a-01a}:
\begin{equation}
{\bf P_{opt}} = \left[ \langle ({\bf P_{\beta_j}^{DM}N)^T P_{\beta_j}^{DM} N} \rangle_{\beta} \right]^{-1}\langle ({\bf P_{\beta_j}^{DM}N)^T P_{\beta_j}^{L}} \rangle_{\beta}
\label{popt}
\end{equation}
where, $ \langle  \rangle_{\beta}$ represents the average over all the $N_{ang}$ directions. {\bf N} is a diagonal matrix of sizes $N_{DM}$x$N_{DM}$ defined by {\bf N}=diag($A_1$,...,$A_{N_{DM}}$), with $A_n$=1 when the considered frequency is smaller than the DM cut-off frequency, and 0 otherwise. ${\bf P_{\beta_j}^{DM}}$ and ${\bf P_{\beta_j}^{L}}$ are constructed in the same way as ${\bf P_{\theta}^{DM}}$ and ${\bf P_{\theta}^{L}}$ in Eq. \ref{eq_propa}. \\
If the mirrors positions exactly match the reconstructed layers, no projection is required and ${\bf P_{opt} = \mbox{Id}}$. In such a situation there is no particular optimization in the FoV, and no additional error due to projection. 
When the mirror positions do not match the turbulent layers, optimizing for a particular FoV position may degrade the correction in other directions. Trade-offs have to be made for a specific set of FoV positions, and the matrix ${\bf P_{opt}}$ performs optimally these trade-offs. In such a situation, an additional error term due to projection arises: the so-called generalized fitting. We discuss this term in Sect. \ref{secpopt}.\\

 \subsection{Derivation of the residual phase power spectrum}
 \label{sect_resPSD}
Putting Eq. \ref{eq_propa}, \ref{measure} and \ref{recons} in Eq. \ref{eqbase}, we then have: 
\begin{equation} 
\tilde{\phi}^{res}_{\theta} = {\bf P_{\theta}^L\tilde{\varphi}_{turb}} - {\bf P_{\theta}^{DM}W(M P_{\alpha}^L\tilde{\varphi}_{turb} + b)}
\label{eqbase2}
\end{equation}

We can now derive the residual PSD for each frequency. By definition, PSD$^{res}_{\theta}$ is the statistical average of the square modulus of the residual phase $\tilde{\phi}^{res}_{\theta}$. Assuming that noise and phase are statistically independent from each other, it follows that: 
\begin{equation}
\begin{array}{l}
\mbox{PSD}^{res}_{\theta} = \\\\
 ({\bf P_{\theta}^L-P_{\theta}^{DM}W M  P_{\alpha}^L)C_{\varphi_n,t}}({\bf P_{\theta}^L-P_{\theta}^{DM}W M P_{\alpha}^L})^T \\\\
+ ({\bf P_{\theta}^{DM}W ) C_{b,t}(P_{\theta}^{DM}W )^T}
\end{array}
\label{eqbaseDSP}
\end{equation}
In the above equation, we have introduced ${\bf C_{\varphi_n,t}}$, the power spectrum matrix of the phases in the turbulent volume and ${\bf C_{b,t}}$ the noise power spectrum matrix. These matrices stand for true noise/atmospheric statistical conditions. They are both diagonal matrices constructed in the same way as ${\bf C_{\varphi_n}}$ and ${\bf C_{b}}$ in Eq. \ref{wtomo}. The differences with ${\bf C_{\varphi_n}}$ and ${\bf C_{b}}$, are that ${\bf C_{\varphi_n,t}}$ can be larger than ${\bf C_{\varphi_n}}$ if we commit a model error, and ${\bf C_{b,t}}$ can have different values of noise variance if we commit a statistical error (see \ref{subsect_modelstat}). We distinguish them from the one introduced as priors in Sect. \ref{sect_mmse} with the subscript $t$. 
In all the following, $\mbox{PSD}^{res}_{\theta}$ will be our working starting point. All the matrices required to construct $\mbox{PSD}^{res}_{\theta}$ are also described in more details in \cite{Neichel-p-08a}.

\section{Simulation Conditions}
\label{simulationcases}
\begin{figure}[h!]
  \begin{center}
            \includegraphics[height=5.5cm]{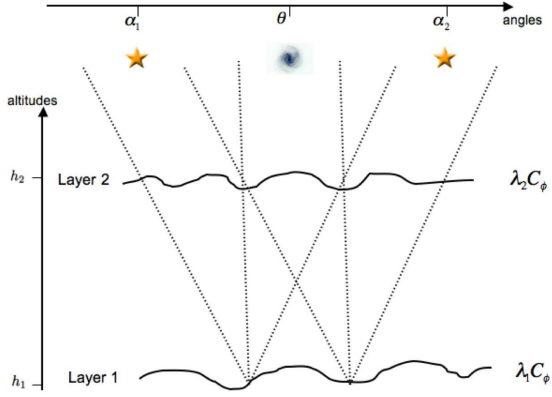}
   \end{center}
  \caption{System geometry for the 2GS two-layers simple case. Altitude of layers is $h_1$=0 and $h_2=h$. Two guide stars are considered in directions $\alpha_1$ and $\alpha_2$, the separation $\alpha_{12}$ between the GSs is 2'. Performance is evaluated at the center of the FoV: $\theta=0$.}
    \label{description2}
   \end{figure}

Based on the residual phase PSD derived in Sect. \ref{sect_resPSD}, we want now to investigate the different errors introduced in Sect. \ref{sec_intro}, respectively: unseen frequencies, model/statistical errors and projection errors. For that we define three simulation cases representative of different GS/FoV configurations:\\
\begin{itemize}
\item The first one, called ``2GS", is a simple 1-dimensional model where the atmosphere is only composed of 2 layers. One of the 2 layers is located in the pupil, the other one is at an altitude h=8km. We consider that the 2 layers have a relative turbulent strength  profile defined by $[\lambda_1,\lambda_2]$. We consider 2 GSs separated by a distance $\alpha_{1,2}$=2arcmin. This simple configuration is mainly used to derive didactic examples, illustrated with the 1-dimensional residual PSD. Fig. \ref{description2} describes the geometry of this case. \\
\item The second case, called ``4GS", is a more realistic model where the atmosphere is composed of 10 layers. The altitudes and strength of the layers are summarized in Table 1. This turbulent volume is sensed with 4GS located on a 2arcmin diameter circle. All GS have the same noise variance $\sigma ^2$=0.5 rd$^2$. This constellation is used as a representative configuration for a medium FoV system.\\
\item The third case, called ``8GS" uses the same turbulence profile, but sensed with 8GS located on a 5arcmin diameter circle. All GS have the same noise variance $\sigma ^2$=0.5 rd$^2$. This constellation is used as a representative configuration for a large FoV system.\\
\end{itemize}

\begin{table}[h!] 
\begin{center}
\caption{10 layer turbulent profile altitudes and relative layer strength.}
\begin{tabular}{|c|c|c|c|c|c|c|c|c|c|c|}
\hline
$h_n$ (m) & 0 & 200 & 600 & 1200 & 3000 & 4700 & 7300 & 8500 & 9700 & 11000 \\ \hline
$\lambda_n$ & 0.41 & 0.16 & 0.1 & 0.09 & 0.08 & 0.05 & 0.045 & 0.035 & 0.02 & 0.01 \\\hline
\end{tabular}
\end{center}
\label{tab2}
\end{table}

Table 2 summarizes the main parameters of these 3 simulation cases. 
\begin{table}[h!]  
\begin{center}
\caption{Simulation configurations.}
\begin{tabular}{|c|c|c|c|}
\hline & \bf{2GS Config.} & \bf{4GS Config.} & \bf{8GS Config.}\\ \hline \hline
Atm. profile & 2lays & 10lays & 10lays \\ \hline
\raisebox{5ex}{Constellation} &  \includegraphics[height=2cm]{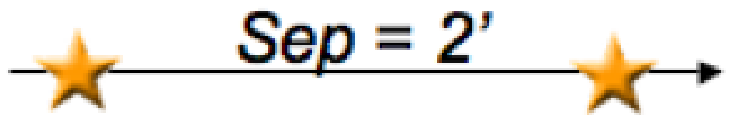}  &  \includegraphics[height=2cm]{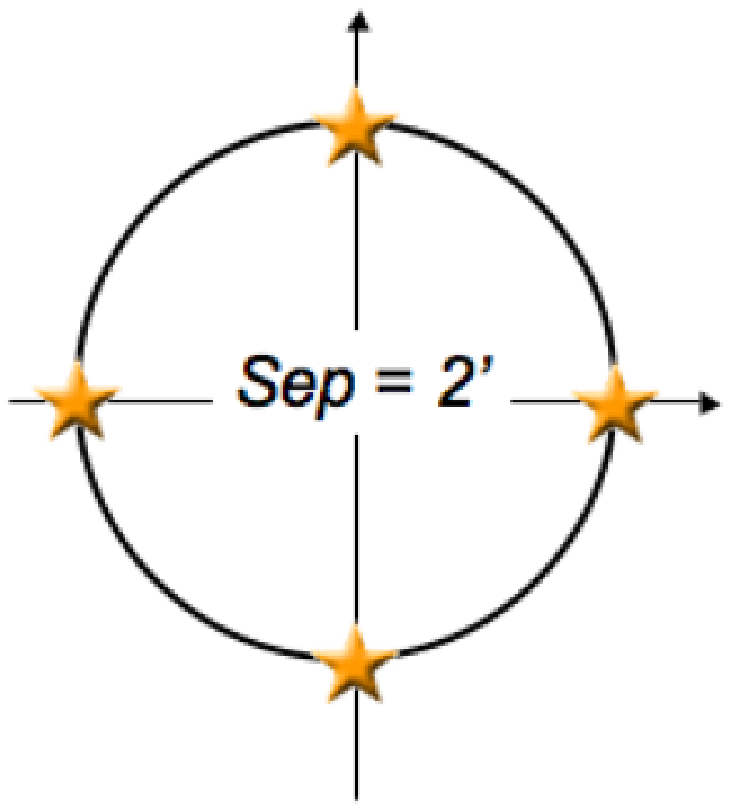}&  \includegraphics[height=2cm]{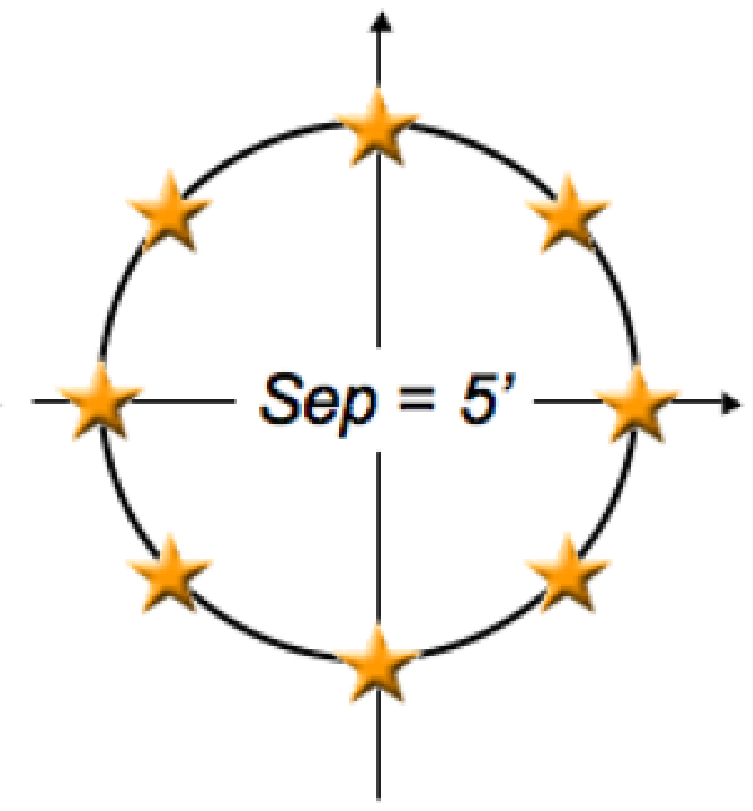}\\ \hline
\# GS & 2& 4 & 8 \\ \hline
FoV Diameter & 2' & 2' & 5' \\ \hline
Noise/GS & 0.5 rd$^2$ & 0.5 rd$^2$ & 0.5 rd$^2$ \\ \hline
\end{tabular}
\end{center}
\label{tab1}
\end{table}

For all simulation cases, the Fried parameter is set to $r_0$=0.12m defined at 0.5$\mu$m and $L_0$=50m. 
We assume the that turbulence measurements are performed with SH WFS. In that case, Rigaut et al. \cite{Rigaut-p-98a} show that the measurement operation, without aliasing and finite exposure time, can be modeled by a $M=2j\pi f_{x,y} \mbox{sinc}(\pi df_{x,y})$ term, where we assumed a square geometry for subapertures with a size (reported in the telescope pupil) equal to $d$. 
For our example, we consider that all WFS have the same subapertures sizes. This implies that all frequencies greater than $f_c = 1/2d$ are not measured. We further assume that all the GSs have the same noise variance defined by $\sigma^2$ = 0.5 rd$^2$, where $\sigma^2$ are the diagonal components of ${\bf C_{b,t}}$. Performance is evaluated at 1.65$\mu m$ (H Band) by projecting the volumetric residual phase in a direction of interest $\theta$. Unless specified otherwise, it is evaluated at the center of the FoV ($\theta$ = 0). \\


In the following sections, we make use of those three simulation cases to investigate (i) pure tomographic reconstruction in Sect. \ref{puretomo}; (ii) model/statistical errors in Sect. \ref{sec_modelerror} and \ref{modelerror}; (iii) projection issues in Sect. \ref{secpopt}. We want to emphasize that most of these errors strongly depend on the turbulence profile. More precisely, we will see that for a given isoplanatic angle, a relevant parameter is defined by the products between layer altitudes and GS separations: $h_n \alpha$. In our analysis, we have chosen to fix the turbulence profile, and investigate these errors in term of GS separation (FoV). Numerical results presented in the following should then not be seen as absolute results, but rather as quantitative insights used to illustrate general trends.\\

\section{Pure Tomographic Reconstruction}
\label{puretomo}
In this section, we consider the ideal situation where noise and turbulent conditions are perfectly known: ${\bf C_{\varphi_n}}$ and ${\bf C_b}$ used in Eq. \ref{wtomo} are then identical to ${\bf C_{\varphi_n,t}}$ and ${\bf C_{b,t}}$ used in Eq. \ref{eqbaseDSP}. This situation is generally called the ``pure tomographic reconstruction".

\subsection{Analytical expressions of the residual PSD}
We first consider the 2GS case. From this simple 1-dimensional configuration, we can derive the analytical form of the residual phase PSD for both filters. Eq. \ref{eqbaseDSP} can be written when considering a (T)LSE as:
\begin{equation}
\left\{
\begin{array}{l}
\mbox{PSD}^{res}_{\theta=0}  = \frac{\sigma^2\left[ 2-cos(2\pi f(\alpha_1 - \theta)h)-cos(2\pi f(\alpha_2 - \theta)h)\right]}{2sin^2(\pi f \alpha_{12}h)M^2} \\
\mbox{ \small{for f}}< f_c \mbox{ \small{ and for non truncated frequencies }} \\\\
\mbox{PSD}^{res}_{\theta=0}  = C_{\phi} \mbox{ \small{for f}}\ge f_c \mbox{\small{  or for truncated frequencies}}
\end{array}
\right.
\label{lsesimple}
\end{equation}

while with an MMSE, it reads:
\begin{equation}
\left\{
\begin{array}{l}
\mbox{PSD}^{res}_{\theta=0} =C_{\phi}  \left\{ \frac{2MC_{\phi} \lambda_1\lambda_2\sigma^2 \left[ 2-cos(2\pi f(\alpha_1- \theta)h)-cos(2\pi f(\alpha_2- \theta)h)\right]+\sigma^4}{4sin^2(\pi f \alpha_{12}h)M^4C_{\phi}^2\lambda_1\lambda_2+2M^2\sigma^2C_{\phi} +\sigma^4}\right\}\mbox{ \small{ for f}}< f_c \\\\
\mbox{PSD}^{res}_{\theta=0}  = C_{\phi}  \mbox{ \small{for f}}\ge f_c
\end{array}
\right.
\label{MMSEsimple}
\end{equation}
\\\\

In Fig. \ref{fig1} we compare these two 1-dimensional residual PSD laws for $\theta=0$, as well as the uncorrected Von-Karman spectrum and the on-axis reconstruction for comparison. The on-axis reconstruction corresponds to a classical LSE Single Conjugated AO configuration and is computed by setting the two GSs in the direction of interest. It follows the typical $f^{-2}$ law expected for derivative WFS \cite{Rigaut-a-92a}. Since the two GSs are on-axis, we benefit of the global flux.\\

\subsection{Notion of unseen frequencies and neutral frequencies}
\label{Sect_notionofunseen}
\begin{figure}[h!]
  \begin{center}
   \begin{tabular}{c}
            \includegraphics[height=6.45cm]{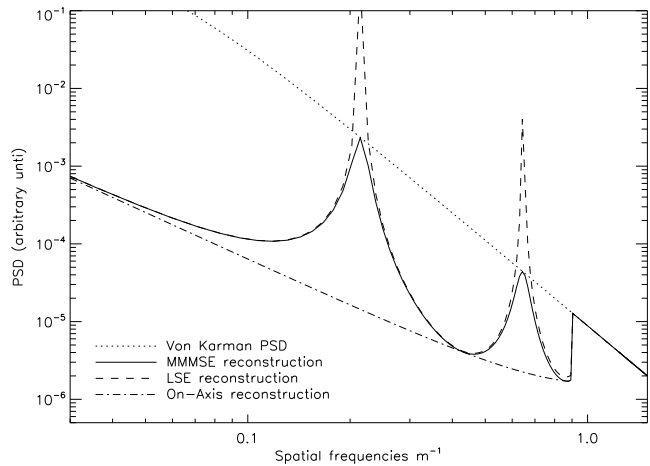}
    \end{tabular}
   \end{center}
 \caption{Comparison of residual PSDs for the MMSE reconstruction (full line) and un-truncated LSE reconstruction (dashed line). Uncorrected Von-Karman (dotted line) and on-axis reconstruction (dashed-dotted line) are also plotted for comparison. In this example, both layers have the same turbulent strength and the WFS cut-off frequency is $f_c=0.9$m$^{-1}$. Two unseen frequencies appear for $f_1$=0.215m$^{-1}$ and $f_2$=0.645m$^{-1}$, two neutral frequencies appear for $f_1$=0.43m$^{-1}$ and $f_2$=0.86m$^{-1}$.}
    \label{fig1}
   \end{figure}

In Fig. \ref{fig1}, it appears that for some specific frequencies, the tomographic residual PSDs (either MMSE or LSE) diverge from the on-axis noise propagation, while other frequencies are following the on-axis noise propagation.\\

To understand the behavior of these PSDs, we first consider the LSE reconstructor.
From Eq. \ref{lsesimple}, we learn that the reconstruction error tends to infinity for each spatial frequency equal to $p/(\alpha_{12}h)$, where $p$ is an integer. Indeed, every time that the period of a frequency exactly match the GS separation, the sum of the phase perturbations induced by the 2 layers is exactly zero, the phase information for these frequencies is lost.
We refer to these frequencies as ``unseen frequencies", as they are not sensed by the WFSs.
For a given turbulence profile, the unseen frequencies only depend on the GS positions: the more distant the GSs, the smaller the spatial frequencies affected, the larger the number of unseen frequencies in a given frequency range of interest. The GS separation can then be understood as a spatial basis which sets the sensitivity of the filter to the different frequencies.\\
We now investigate what is the impact of the unseen frequencies when looking in a particular direction of interest $\theta$.
This relies on the numerator of Eq. \ref{lsesimple}. We find that for some specific frequencies this numerator goes to zero. This happens each time that the sum of the phase perturbations induced by the 2 layers is exactly zero in the direction $\theta$. 
These frequencies are not impacting the performance in the direction of interest, we will called them ``neutral frequencies".
For instance when the direction of interest is at $\theta$=0 (as in Fig. \ref{fig1}), these neutral frequencies occurs each $f = 2p/(\alpha_{12}h)$. Therefore, for this specific direction, the neutral frequencies are canceling one unseen frequency over two.
We note that for the frequency exactly corresponding both to an unseen and neutral frequency, Eq.\ref{lsesimple} is undetermined. For that specific frequency, calculation of the limits gives:
\begin{equation}
\mbox{PSD}^{res}_{\theta=0} =\left( \frac{\sigma^2}{2M^2}\right)
\end{equation}
which is the classical on-axis propagation law \cite{Rigaut-p-98a}. \\
Another example of the effect of the neutral frequencies is when the direction of interest is exactly lying on one of the GS (e.g. $\theta = \alpha_1$). In such a situation all the unseen frequencies are canceled by the neutral frequencies: an unseen frequency has no effect on the image quality in the GS directions.\\

\begin{figure}[h!]
  \begin{center}
   \begin{tabular}{c}
            \includegraphics[height=6.45cm]{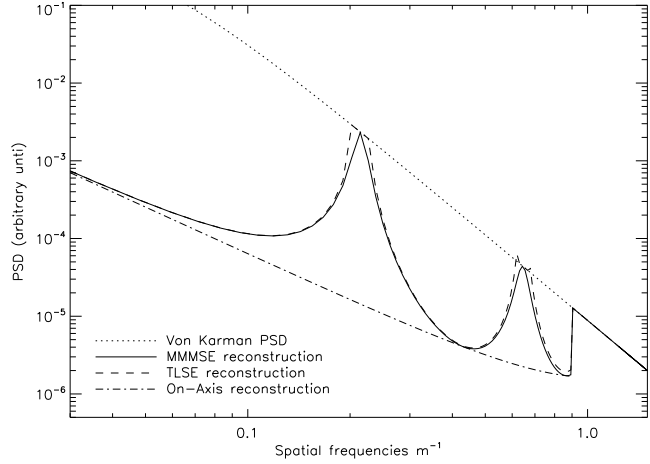} \\
          \end{tabular}
   \end{center}
  \caption{Same as Fig.\ref{fig1} with TLSE reconstructor. Threshold was optimize to minimize the residual variance.}
   \label{filters}
   \end{figure}

For the pure LSE reconstructor, unseen frequencies are a real issue as they lead to an over-amplification of noise. For these frequencies, the interaction matrix coefficients go to zero, direct invert is ill-conditionned and the noise is dramatically amplified. 
Around an unseen frequency, there is a set of badly-seen frequencies. For these frequencies, measurement is close to zero, but since the WFSs are noisy, the same problem as for the unseen frequencies applies. In the following, we will also refer to these frequencies as unseen frequencies.\\
To avoid this noise amplification, one should use the TLSE reconstruction.
In Fig. \ref{filters} we show the corresponding residual PSD for the TLSE reconstructor having the best threshold. Noise amplification is avoided, at the price of uncorrected frequencies. \\
   
\begin{figure}[h!]
  \begin{center}
   \begin{tabular}{c}
            \includegraphics[height=6.45cm]{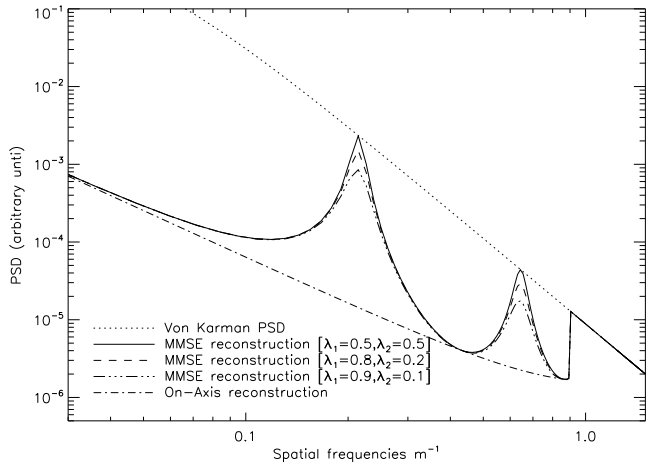} 
          \end{tabular}
   \end{center}
   \caption{Influence of turbulence strength repartition on MMSE reconstruction error.}
    \label{Cn2}
   \end{figure}

Contrary to the TLSE reconstructor, the MMSE reconstructor does not require any truncation. In fact, the MMSE filter includes prior knowledge of the SNR by the mean of the regularization term ${\bf \sigma^2 C_{\varphi_n}^{-1}}$ (see Eq. \ref{eq_regul}). 
This regularization appears as an additive term in the numerator of Eq. \ref{MMSEsimple} which weight the noise propagation.
Thanks to this regularization term, the MMSE is able to optimize the reconstruction depending on the SNR: for frequencies that have a good SNR, MMSE and TLSE are equivalent, for poor signal to noise frequencies, the invert of the interaction matrix is weighted by the regularization term, avoiding noise amplification. It follows that, (i) even on a pure unseen frequency, the residual MMSE PSD is never higher than the $C_{\phi}$ signal variance, and (ii) whatever the truncation threshold used for the TLSE estimator, the MMSE approach always gives smaller residual variance (see also \cite{Fusco-a-01a}).\\
Following with the MMSE, it is interesting to note that for a frequency exactly on an unseen mode, the residual PSD writes as:
 \begin{equation}
\mbox{PSD}^{res}_{\theta=0}  = C_{\phi}  \left( \frac{8M^2C_{\phi} \lambda_1\lambda_2+\sigma^2}{2M^2C_{\phi} +\sigma^2}\right)
\label{eq_MMSEonunseen}
\end{equation}
Eq. \ref{eq_MMSEonunseen} indicates that when the turbulence is equally distributed between the two layers ($\lambda_1$=$\lambda_2$=0.5),  the residual PSD exactly reaches the uncorrected spectrum. However, as soon as the turbulent strength profile is not uniformly distributed, the residual error is always lower than the uncorrected spectrum. We illustrate this behavior in Fig. \ref{Cn2} with three residual PSDs computed for respectively (from top to bottom) [$\lambda_{1}$=0.5,$\lambda_{2}$=0.5], [$\lambda_{1}$=0.7,$\lambda_{2}$=0.3] and [$\lambda_{1}$=0.9,$\lambda_{2}$=0.1] (the [$\lambda_{1}$=1,$\lambda_{2}$=0] case is superimposed with the on-axis reconstruction error). 
This behavior can be explained by the fact that when the turbulence is equally distributed, the chance that the sum of both layers gives a null measurement is maximal. Statistically, no information can be extracted from the measurement.
When the two layers do not have the same strength, this probability decreases and even for an unseen frequency, some signal can be measured. As a limiting configuration, if one layer is free of turbulence ($\lambda_{1,2}=0$), we are able thanks to this prior information to entirely recover the incoming perturbation. As the MMSE filter includes the knowledge of turbulence strength per layer, it is able to discriminate these conditions, and to modulate the regularization term layer per layer. This also illustrates how the prior knowledge of the statistical conditions can consequently reduce the residual variance compared to the crude TLSE approach.

\subsection{Generalization to realistic cases}
\begin{figure}[h!]
  \begin{center}
   \begin{tabular}{l}
            \includegraphics[height=3.5cm]{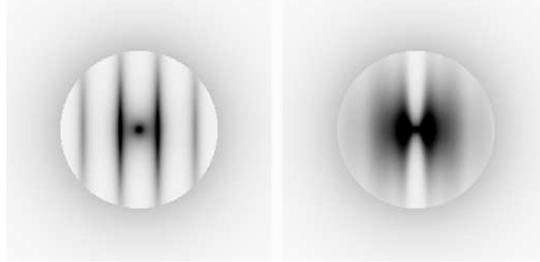}
         \end{tabular}
   \end{center}
   \caption{2-dimensional residual PSDs for the 2GS constellation with the 2 layers profile (left) and the 10 layer profile (right). Log scale and inverted colors are used. The GS orientation is the same as in Table 2.}
    \label{DSP2GS}
   \end{figure}

To understand what is the impact of a larger number of layers/GSs on unseen frequencies, we first start with a generalization of the simple 2GS case. In Fig. \ref{DSP2GS} we show the residual PSD for the 2GS case, but treated as a 2-dimensional configuration. On the left, we keep the two layers atmosphere, whereas on the right we use the 10 layer profile as defined in Table 1. We only display here the results of the MMSE reconstructor. Note that a cut along the $x$ coordinate on the left PSD of Fig. \ref{DSP2GS} would give the same MMSE reconstructor plot as the one presented in Fig.\ref{fig1}.\\

\begin{figure}[h!]
  \begin{center}
   \begin{tabular}{l}
            \includegraphics[height=5.5cm]{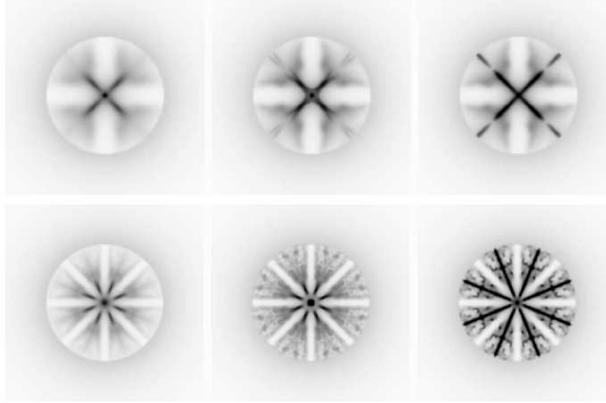}
         \end{tabular}
   \end{center}
   \caption{2-dimensional residual PSDs for the 4GS (top) and 8GS (bottom) cases. From left to right: MMSE, TLSE, non truncated LSE. Log scale and inverted colors are used. All PSDs are displayed with same scale. The GSs orientation are the same orientation as in Table 2.}
    \label{DSP10lays}
   \end{figure}

If we compare the shape of the two PSD presented in Fig. \ref{DSP2GS}, we easily understand what is the impact of multiplying the number of layers: unseen frequencies become sectors instead of localized energy peak. Indeed, with several layers, each pair of turbulent layer creates is own set of unseen frequencies, so the energy is spread upon more frequencies.\\
The second important point to emphasize is that unseen frequencies only appear in the parallel direction of the GS geometry. As explained in Tokovinin \& Viard  \cite{Tokovinin-a-01a}, for all the frequencies perpendicular to the GS direction ($f$ = \{$f_x$=0;$f_y$\} for the 2GS case), the effective separation between the GSs in each layer ($\alpha_{1,2} h_n$) will not change the measured phase of those  frequencies. These frequencies are always perfectly measured, the noise propagation is the one of the classical AO situation.
This explains the ``clean" frequency area along the $y$ coordinate in Fig.\ref{DSP2GS}. The width of this ``clean" frequency area only depends on the GS separation that dictates the lowest unseen frequency.\\
We are now able to understand the generalization to the 4GS-8GS cases.
In Fig. \ref{DSP10lays}, we show the 2-dimensional residual PSDs for the MMSE (left), the TLSE (center), and the LSE (right) for respectively the 4GS (top) and 8GS (bottom) constellations.\\

The 4GS case is nothing else than the product of two 2GS constellation rotated by 90$^{\circ}$. We retrieve two ``clean" frequency area oriented toward the GSs directions. The same argument can be applied to the 8GS constellation. In addition, the diameter of the 8GS constellation is larger than the 4GS one: the width of its  ``clean" frequency area is smaller.\\

In Fig. \ref{DSP10lays}, it also appears that the variance is much higher for LSE or TLSE reconstructors than for the optimal MMSE one as already explained for the 2GS case in Fig. \ref{filters}.\\

Finally, we note that as expected the number of low frequencies affected by unseen modes is larger for the large FoV configuration that is the 8GS case than in the 4GS configuration. The noise level associated to each frequency is however lower because twice more GSs are used. It results that both the 4GS and the 8GS configuration give similar residual variance for the MMSE reconstruction: $\sim$0.77 rd$^2$ (with our simulation conditions).

\section{Model Errors}
\label{sec_modelerror}
Up to now, we have assumed that the priors were perfectly tuned, i.e., that they do correspond to real conditions. These priors comes from our knowledge of the system/atmospheric conditions, and are not always easily accessible. Moreover, they can evolve during an observation run and it could be impractical to re-compute the reconstruction matrix frequently. In any case, it is of prime importance to understand and quantify the impact of these statistical/model errors on the reconstruction process. \\
Still using the three configurations presented in sect. \ref{simulationcases}, we first investigate the model errors with an error on the altitudes of the reconstructed layers and an error on the number of reconstructed layers. These errors occur both in the TLSE and the MMSE reconstructors. We then aim to (i) understand the impact of the model errors illustrated with the 2GS case; (ii) compare the relative performance and robustness of the TLSE and MMSE reconstructors with the 4GS configuration; (iii) compare the medium (4GS) and large FoV (8GS) systems in the MMSE approach. \\

\subsection{Error on layer altitude}
\label{layeraltitude}
\begin{figure}[h!]
  \begin{center}
   \begin{tabular}{c}
            \includegraphics[height=13cm]{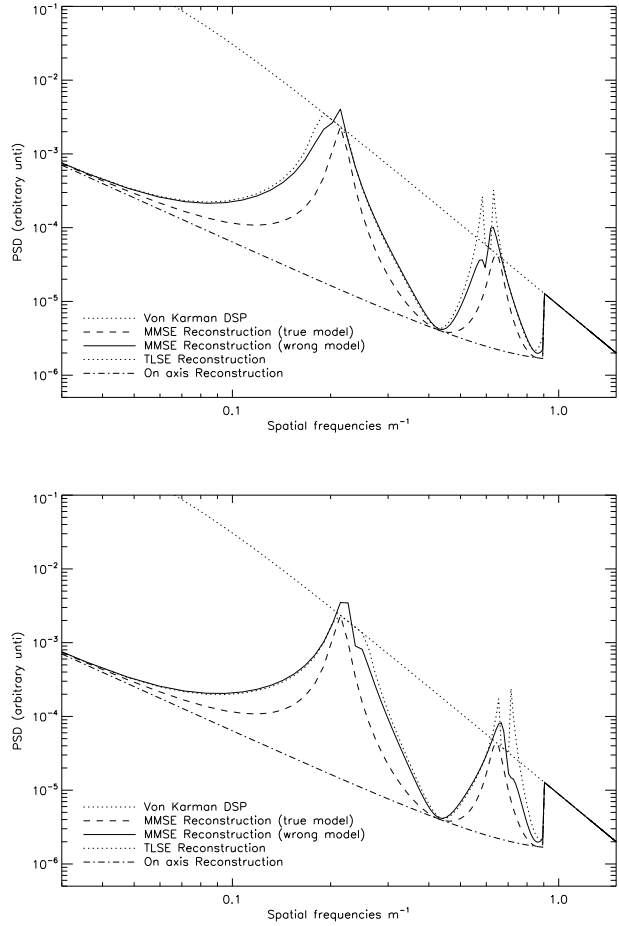} 
          \end{tabular}
   \end{center}
   \caption{Residual PSD for an error in the altitude of the reconstructed layer. {\bf Top:} Upper altitude is overestimated (8.5km instead of 8km). {\bf Bottom:} Upper altitude is underestimated (7.5km instead of 8km). }
    \label{altitude1}
   \end{figure}

We first start with the 1-dimensional 2GS case. We recall that only two layers are considered in this case. In Fig.\ref{altitude1} we show an example for which (i) the upper reconstructed layer is higher than real layer (Top panel) and (ii) the upper reconstructed layer is lower than real layer altitude (Bottom panel). In both situations, the residual PSD is strongly affected, and for some frequencies the residual variance rises above the uncorrected PSD.\\
An error on the altitude where the layers are reconstructed affects the frequencies where unseen modes should be regularized/truncated. In fact, the altitudes of the reconstructed layers introduced in the model set the sensitivity of the reconstructors to unseen frequencies. A wrong geometry implies that wrong modes are going to be regularized/truncated. For the frequencies that correspond to well-seen frequencies for the model, the direct invert of the interaction matrix is performed, whereas these frequencies could correspond to badly seen modes in the real geometry. For theses frequencies, an over amplification of noise could appear. On the opposite, some frequencies that are well seen by the real geometry could be treated as unseen frequencies by the model. For these modes, whereas a direct inverse would have worked well, over regularization or truncation limits the accuracy of the reconstruction. 

\begin{figure}[h!]
  \begin{center}
   \begin{tabular}{c}
          \includegraphics[height=6.45cm]{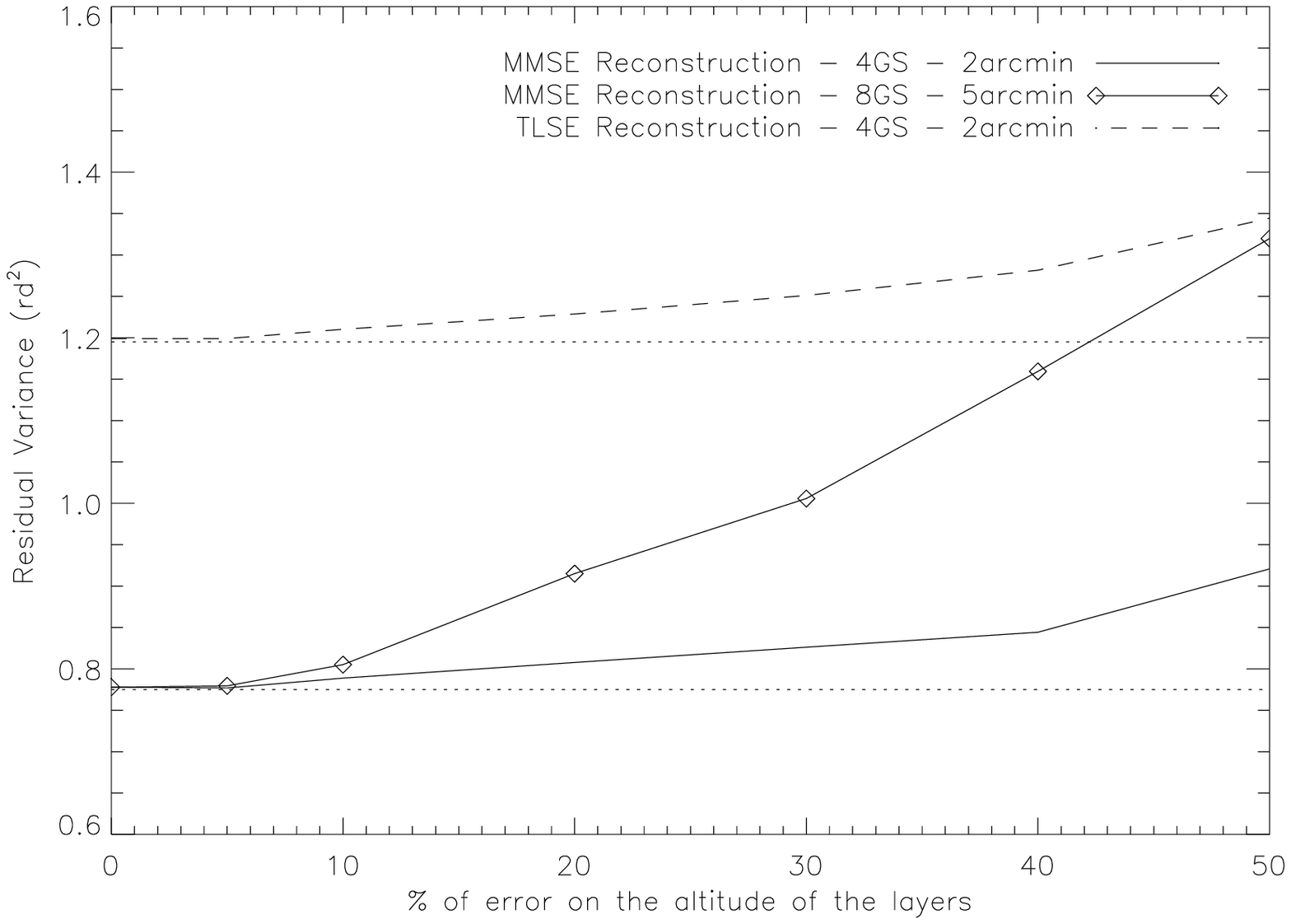} 
          \end{tabular}
   \end{center}
   \caption{10 layers profile. Influence of an error on reconstructed altitudes of each layer. \% of error means that each reconstructed layer is at an altitude of H$\pm$H*X\%. }
    \label{Altitude3}
   \end{figure}

In a more realistic configuration, the atmosphere includes several dominant layers, and this error will be minimized. Of course, the best match between real profile and dominant layers will give the best results. 
To illustrate this point, we investigate our realistic 10 layer cases (4GS-8GS). We reconstruct the 10 layers, but for each reconstructed layer we allow an error in altitude of $\pm$X\%, X going from 0\% to 50\%. For each altitude error, we perform 50 random trials and we compute the mean residual variance. Results are shown in Fig. \ref{Altitude3}. \\

We first compare the results obtained for the 4GS case. We find that both reconstructors (MMSE and TLSE) follow the same tendency: the residual variance increases with the error on the model altitude. In addition, the order of magnitude of this error is more or less the same for both reconstructors.\\
If we now compare the results of the 4GS and 8GS cases, we notice that the large FoV 8GS configuration is much more sensitive to an error in layer altitudes than the medium FoV 4GS one. For the medium FoV configuration, the impact of this error is moderated: an error of 50\% on the layer altitudes increases the residual variance by $\sim$20\% with our turbulence profile. This indicates that a perfect knowledge of layer altitude is not necessary, as soon as many layers are reconstructed. However, this is no more true for the large FoV configuration: an error of 50\% increases the residual variance by almost a factor of 2. As an analogy to what is explained by R. Ragazzoni \cite{Ragazzoni-a-01a}, the larger the distance between reconstructed and true layer, the lower the maximal equivalent frequency reconstructed in the true layer. And this maximal equivalent cut-off frequency depends on the FoV. 
This can also be explained with Fig. \ref{DSP10lays}: when the FoV is increased, the unseen frequencies cover a larger area and the  noise amplification affects more and more frequencies.\\
We conclude that this error term is impacting the tomographic reconstruction for large FoV configurations, and that either a good knowledge of the turbulence profile or the use of more GS to reduce the unseen frequencies area is required to limit the impact of this model error.

\subsection{Error on number of layers}
\label{nboflayer}
\begin{figure}[h!]
  \begin{center}
   \begin{tabular}{c}
          \includegraphics[height=6.45cm]{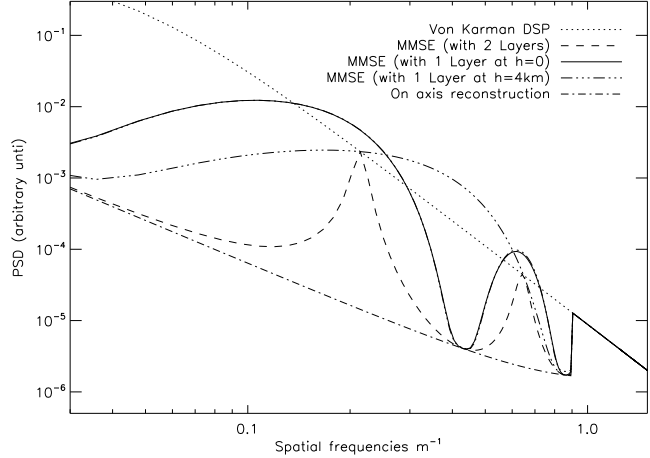} 
          \end{tabular}
   \end{center}
   \caption{Residual PSD for a model error in the number of layers. The real profile is made of  two-layers, whereas the reconstruction model use only one layer either in the telescope pupil (full line) or at an altitude of 4km (dashed dotted line). The MMSE and TLSE cases are superimposed.}
    \label{nbLays}
   \end{figure}

For computing reasons, the reconstruction is performed on a limited number of layers, generally smaller than real turbulence conditions. In that configuration, the reconstructed volume does not match with real profile, and the resulting error can be important. 
We first illustrate the impact of an incorrect number of reconstructed layers with the 2GS case. As only 2 layers are used to model the real turbulence profile, a unique layer is used in the reconstruction. In Fig. \ref{nbLays}, we illustrate this situation when the reconstructed layer is in the telescope pupil (full line) and when it is at an altitude of 4km (dashed-dotted line). It appears that the residual PSD is strongly influenced by this model error: the domain of frequency affected by noise amplification is significantly enlarged. Interestingly, when the reconstructed layer falls just between the two real layers, this noise amplification is reduced compared to a reconstruction in the telescope pupil. Indeed, we can show that if the real turbulence is equally distributed between the two layers, the optimal altitude for the reconstructed layer is just at the middle. If the real turbulence is not equally distributed, the optimal altitude for reconstruction moves toward the strongest layer.

\begin{figure}[h!]
  \begin{center}
   \begin{tabular}{c}
          \includegraphics[height=6.45cm]{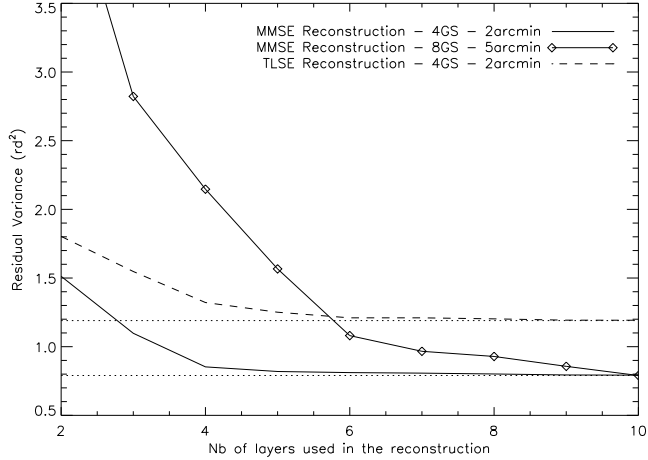} 
          \end{tabular}
   \end{center}
   \caption{Influence of the number of layers used in the reconstruction process. Real profile is the 10 layers.}
    \label{nlays}
   \end{figure}

We now investigate the effect of this model error in presence of a ``real" turbulence profile. To do so, we use the 4GS/8GS case, the ``real" turbulence profile is defined by the 10 layers one, and we progressively increase the number of reconstructed layers.
This has been introduced by Fusco \cite{Fusco-a-99a} as Equivalent Layers (EL).
We then follow their approach for the definition of the altitudes/strenght of the EL: the true profile is divided into $N_{el}$ regularly spaced slabs, for each slab an equivalent height and strength is computed.
Results are shown in Fig. \ref{nlays}. 
This figure shows that for the medium FoV constellation (4GS), a reconstruction on only few layers (typically 3 or 4) is enough to reduce consequently the remaining error. With the 4GS case, we also find that the MMSE residual variance converge more rapidly than the TLSE one. For the large FoV configuration (8GS), we observe that at least 8 layers must be reconstructed, and this number would certainly be larger if the initial profile had included more layers. As for an error on layer altitudes, the tomographic reconstruction is very sensitive to this model error, and particularly when the size of the FoV increases. We draw similar conclusions as in Sect. \ref{layeraltitude}: if the number of GS is limited, a good knowledge of turbulence profile is necessary for an accurate reconstruction.

\section{Statistical Errors}
\label{modelerror}

We have seen in the previous section the effect of the model errors, we now investigate the impact of the statistical errors. 
We then assume that the model is perfectly known: we always reconstruct the real number/altitudes of layers. Instead, we explore the consequences of (i) an error on the noise level (${\bf C_b} \ne {\bf C_{b,t}}$), (ii) an error on the global strength of turbulence ($r_0 \ne r_{0,t}$), and (iii) an error on the relative repartition of the turbulence ($\lambda_n \ne \lambda_{n,t}$).\\

An error on these statistical parameters will impact on the value of the SNR used by the model (Cf. Eq. \ref{eq_regul}). Basically, the SNR used by the model sets the number of filtered/truncated frequencies.
If this SNR is overestimated, the number of filtered/truncated frequencies decreases. For unseen frequencies, the model assumes that the SNR is good, a direct invert of the interaction matrix is performed, and then we expect an over amplification of the noise localized on unseen frequencies. On the other hand, if the SNR is underestimated, the weight of the regularization term is increased: too much frequencies are filtered/truncated whereas they could have been corrected. We then expect a broadening of the PSD around the unseen frequencies, but no noise over-amplification.

\subsection{Error on noise priors}
\begin{figure}[h!]
  \begin{center}
   \begin{tabular}{c}
            \includegraphics[height=13cm]{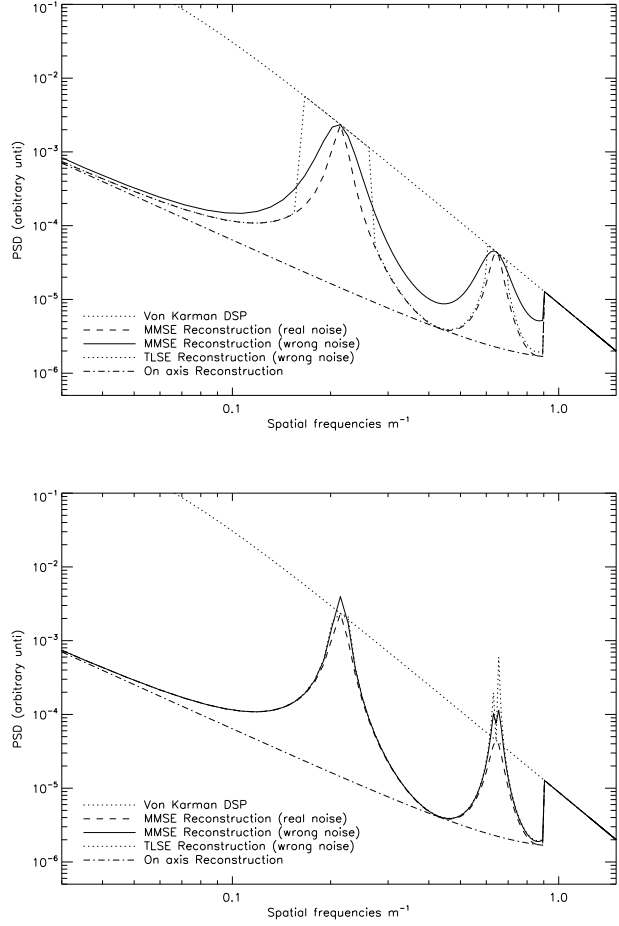}
          \end{tabular}
   \end{center}
 \caption{Residual phase PSD for an error in noise priors. {\bf Top:} Overestimation of noise by a factor of 10: ${\bf C_b} = 10{\bf C_{b,t}}$. {\bf Bottom:} Underestimation of noise a factor of 10: ${\bf C_b} = {\bf C_{b,t}}/10$.}
   \label{noise}
   \end{figure}

In Fig. \ref{noise}, we first use the 2GS case to illustrate the effect of an error on the noise estimation.
The top plot in Fig. \ref{noise} shows the impact of an overestimation of the noise variance by a factor of 10 (${\bf C_b} = 10{\bf C_{b,t}}$), and the bottom plot the impact of an underestimation of the noise variance by a factor of 10 (${\bf C_b} = {\bf C_{b,t}}/10$).
With this simple example, we indeed find that if the noise is overestimated (top figure), the residual PSD tends too rapidly to the uncorrected spectrum and too much frequencies are filtered/truncated. If the noise is underestimated (bottom figure), an over amplification of the residual error appears on the unseen frequencies and not enough frequencies are filtered/truncated.\\

\begin{figure}[h!]
  \begin{center}
   \begin{tabular}{cc}
            \includegraphics[height=6.45cm]{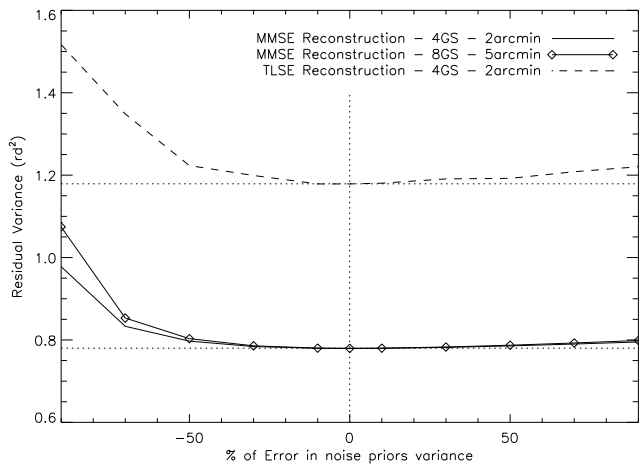} 
          \end{tabular}
   \end{center}
 \caption{Residual variance of the MMSE(full-line) and TLSE (dashed-line) reconstructors with wrong noise model. Error in noise are given in \% of real noise. Dotted-lines symbolize the minimal variance level when no errors on noise priors are committed.}
   \label{noise4}
   \end{figure}

Although the residual PSDs are affected by a wrong noise model, at first glance, the impact of an error in noise statistics does not seem to increase significantly the residual error. 
To quantify this effect, we use our realistic cases (4GS-8GS) and we plot in Fig. \ref{noise4} the residual variance for a MMSE/TLSE reconstructor having wrong noise priors. The error in noise priors is given in \% of real noise: for instance, $\pm$50\% of error in noise variance corresponds to ${\bf C_b} = \pm1.5{\bf C_{b,t}}$.
From Fig. \ref{noise4}, we learn that: 
\begin{itemize}
\item the residual variance in the MMSE approach is always smaller than the TLSE one, which means that even with wrong priors, the MMSE reconstructor gives better results than the TLSE one.
\item The TLSE reconstructor is slightly more sensitive to noise error than the MMSE one.
\item The large FoV configuration (8GS) is slightly more sensitive than the medium FoV one (4GS).
\item For both reconstructors, if the noise level is not perfectly known, a conservative approach would be to overestimate the noise  priors.
\end{itemize}

\begin{figure}[h!]
  \begin{center}
   \begin{tabular}{cc}
            \includegraphics[height=6.45cm]{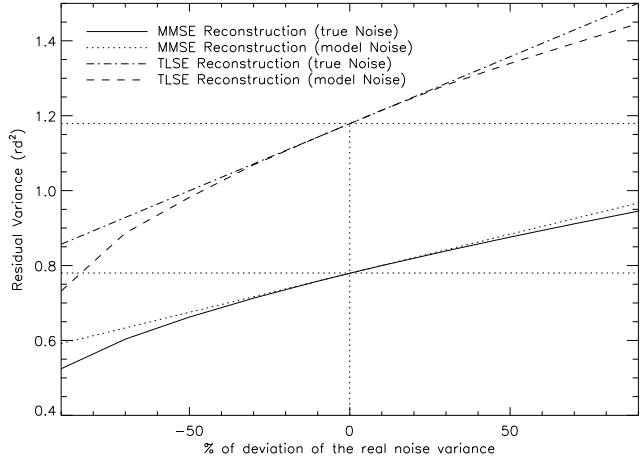} 
          \end{tabular}
   \end{center}
 \caption{Robustness of the MMSE/TLSE reconstructor to changes of ``real" conditions. The ``real" noise level (defined by ${\bf C_{b,t}}$) ranges from -90\% to +90\% around a reference level (0.5 rd$^2$ when ``\% of deviation of the real noise variance" = 0). The full and dashed lines show the residual variance for perfectly tuned MMSE and TLSE reconstructors: noise priors follow the real noise variations. Dotted and dotted-dashed lines show the residual variance when noise priors are set to the reference noise level.}
   \label{noise_example}
   \end{figure}

Another way to test the robustness of these reconstructors to the noise statistical error, is to study for a given reconstructor computed with a reference noise level, the impact on the performance when real conditions are changing. This is illustrated in Fig. \ref{noise_example} for the 4GS configuration. Both MMSE and TLSE reconstructors are computed for a given reference noise level (``variation for real noise variance" equal 0), and we compute the residual variances when real noise conditions are changing. The change in real noise variance  is given in \% of noise priors: for instance, $\pm$50\% of variation in real noise variance corresponds to ${\bf C_{b,t}} = \pm1.5{\bf C_{b}}$. In that situation, we find that the variation of the performance due to real conditions is much more impacting the residual variance than the noise model error itself. Indeed, an increase of the real noise variance by 50\% leads to an increase of the residual variance by $\sim$10\%, whereas the corresponding model error only increases the residual variance by $\sim$3\% with our parameter set. Results for the 8GS configuration are very similar. 

Combining the results of Fig. \ref{noise4} and Fig. \ref{noise_example}, we conclude that a perfect knowledge of the noise priors is not required to obtain an accurate tomographic phase reconstruction.

\subsection{Error on turbulent strength priors}
\subsubsection{error on $r_0$}
The global strength of the turbulence $r_0$ is included in the estimation of the SNR. The conclusions drawn in the previous section are then exactly transposable here: (i) the MMSE is always better than the TLSE with wrong truncation threshold (ii), the large FoV configuration is slightly more sensitive than the medium FoV one, and (iii) a conservative approach would be to underestimate the global strength of the turbulence. \\
 
\subsubsection{error on turbulence strength per layer}
\begin{figure}[h!]
  \begin{center}
   \begin{tabular}{c}
            \includegraphics[height=13cm]{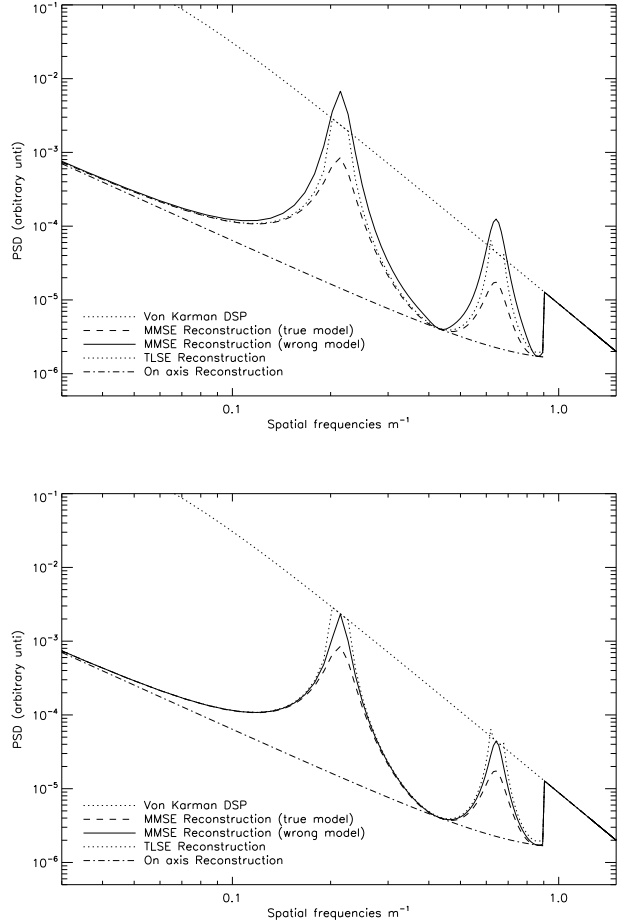} 
          \end{tabular}
   \end{center}
   \caption{Error on turbulence strength priors with real turbulence distribution defined by [$\lambda_{1,t}$=0.9, $\lambda_{2,t}$=0.1] . {\bf Top:} turbulence strength repartition used by the model is [$\lambda_1$=0.1,$\lambda_2$=0.9] (solid line). For comparison, we plot the residual PSD if the model were correct (dashed line). {\bf Bottom:} turbulence strength repartition used by the model is [$\lambda_1$=0.5,$\lambda_2$=0.5].}
    \label{Cn22}
   \end{figure}

The turbulence strength distribution is another essential parameter to be introduced in the regularization process. This term only affects the MMSE reconstructor, as the TLSE one does not include the profile distribution information. 
We have seen in Sect. \ref{Sect_notionofunseen} that the turbulence strength distribution ($\lambda_n$) sets the SNR per layer.
An error on the repartition of the turbulence strength will then produce an error on the SNR, and an over amplification of noise.\\
We illustrate this point in Fig. \ref{Cn22} (top) with a pessimistic example: the turbulence strength repartition used by the model is [$\lambda_1$=0.1,$\lambda_2$=0.9], whereas real turbulence distribution is [$\lambda_{1,t}$=0.9, $\lambda_{2,t}$=0.1].
For this pessimistic situation, we see in Fig. \ref{Cn22} that the residual PSD rises above the uncorrected turbulent PSD. 

To avoid any noise amplification, a conservative approach would be to feed the model with a uniform turbulent strength distribution.
This example is illustrated in Fig. \ref{Cn22} (bottom). The true turbulent distribution is still [$\lambda_{1,t}$=0.9, $\lambda_{2,t}$=0.1], the one used by the model is [$\lambda_1$=0.5,$\lambda_2$=0.5].
With such a model, we have seen with Eq. \ref{eq_MMSEonunseen} that the residual PSD becomes independent of the modeled strength repartition. As expected, the constant profile is the one providing the less information.\\
With the conditions of  Fig. \ref{Cn22} (top), the total variance of the MMSE is higher than the TLSE one. However, following the conservative approach, i.e., the constant profile (Fig. \ref{Cn22} (bottom)), the MMSE is always better than the best tuned TLSE. \\\\
We could further investigate these model/statistical errors, for instance by combining several terms together. In that case, it is interesting to show that some model errors can be compensated from others (e.g. \cite{Petit-a-08a}). For instance, the global variance due to a mis-knowledge of layer altitude could be compensated by using more noise priors in the reconstruction.
A pure quantitative study of model errors depends on system characteristics, and this exhaustive and specific work is certainly out of the scope of this paper. Instead, in the next section we choose to investigate the error due to the projection onto DMs.

\section{Projection onto DMs}
\label{secpopt}
\begin{figure}[h!]
  \begin{center}
   \begin{tabular}{c}
          \includegraphics[height=13cm]{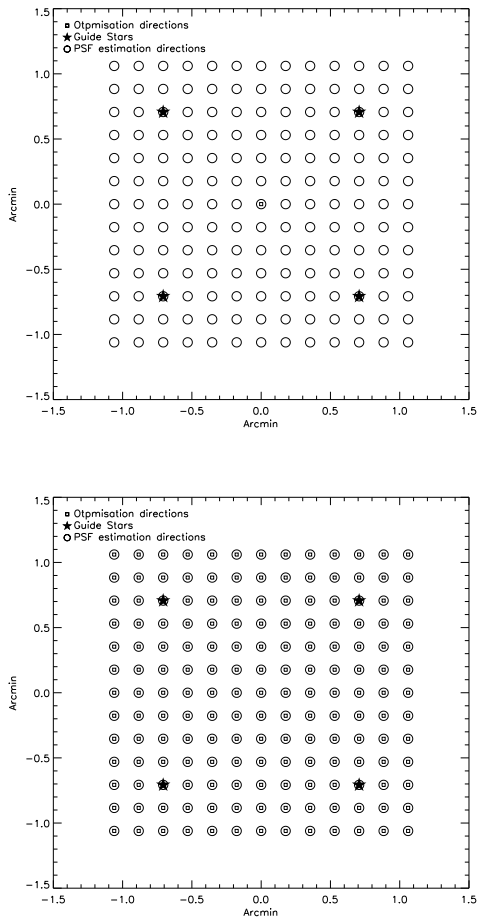} 
          \end{tabular}
   \end{center}
   \caption{Geometry used in the simulations. {\bf Top:} One direction of optimization is considered at the center of the field. {\bf Bottom:} optimization is performed all over the field.}
    \label{config_optimi}
   \end{figure}

Up to now, we have only considered the tomographic phase reconstruction issue. In this section, we now investigate the projection onto deformable mirrors, and the supplementary error term related to generalized fitting. This term relies on Eq. \ref{popt} and depends on the optimization positions (\{$\beta_j$\}) and the number/altitudes of DMs.  
We mainly use the 4GS constellation, and we investigate different WFAO systems for a 42m telescope.
We first consider a projection onto a single DM with one direction of optimization (MOAO/LTAO), we then increase the optimization field still assuming a single DM (GLAO), and finally we introduce several DMs with an optimization in the field (MCAO). \\
Complementary to what have been done in the previous sections, we now extend the analysis to the whole field instead of concentrating on a particular direction. For each configuration, we choose to evaluate the performance in term of Strehl Ratio (SR), in 169 directions regularly spaced on a 13x13 grid across the field. Note that the SR metric is particularly relevant for diffraction-limited imagery (MOAO/LTAO/MCAO), but less appropriate for partial atmospheric corrections (GLAO, e.g. \cite{Tokovinin-p-08a}). Nevertheless, for the sake of clarity and simplicity, we have decided to only consider SR in the following. \\
For MOAO, optimization is done at the center of the field, for GLAO/MCAO optimization is done on the same directions. Fig. \ref{config_optimi} illustrates the geometry used for the MOAO/LTAO configuration (top) and for the GLAO/MCAO configurations (bottom). Note that the GS constellation is rotated by 45$^{\circ}$ compared to Table 2.\\

In all the following, except in Sect.\ref{sec_equivalentlays}, no model/statitical errors are introduced. We further assume that DM pitchs match WFS pitchs, which would correspond to DMs with $\sim$75x75 actuators. Atmospheric parameters are the ones introduced in Sect. \ref{simulationcases}. We stress that a different set of atmospheric parameters would lead to different numerical results. The derived performance is then only indicative and useful for a relative comparison of the systems. Finally, we recall that the telescope diameter is set to 42m.

\subsection{MOAO/LTAO: No projection error}
For some specific observations a uniform correction of the whole FoV is not required. For instance, in 3D spectroscopy extra-galactic studies, only few directions of interest must be corrected for. This can be achieved by the way of MOAO or LTAO (\cite{Hammer-p-02a, Assemat-a-07a}). In these concepts, several GSs spread over the field are used to perform the tomographic reconstruction of the turbulent volume. The correction is then applied with one DM conjugated to the pupil per science channel: ${\bf P_{\theta}^{DM} = Id}$. For LTAO, only one science channel is considered, for MOAO, few channels are corrected at the same time. For each direction of optimization $\beta$, the projection onto the DM simply writes as:
\begin{equation}
{\bf P_{opt}} = {\bf P_{\beta}^L}
\end{equation}
In those directions, the performance is the one of the pure tomographic reconstruction described in Sect. \ref{puretomo}, there is no supplementary term of error due to generalized fitting. An example of the expected performance of an LTAO/MOAO system is presented in Fig. \ref{moaoglao} (top) for the 4GS case. In this example, only one direction at the center of the field ($\beta$=0) has been optimized. For this direction, the performance is only limited by unseen frequencies and it reaches $\sim$50\% of SR. Outside this direction, performance quickly decreases due to classical anisoplanatism.\\

\subsection{GLAO}
\begin{figure}[h!]
  \begin{center}
   \begin{tabular}{c}
          \includegraphics[height=11cm]{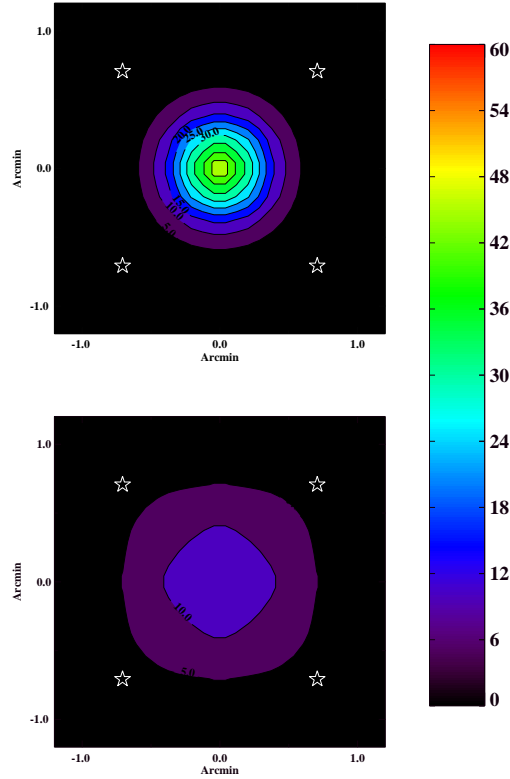} 
          \end{tabular}
   \end{center}
   \caption{SR maps for MOAO (top) and GLAO (bottom). GS are 45$^{\circ}$ rotated compared to Table 2. The SR have been linearly interpolated on a 26x26 grid. For MOAO, the SR is $\sim$50\% at the center of the field. For GLAO, the SR is $\sim$10\% at the center of the field.}
    \label{moaoglao}
   \end{figure}

With still one DM conjugated to the telescope pupil, but an optimization in the whole field, we now investigate GLAO \cite{Rigaut-p-02a}.
The GLAO correction represents the worst case for generalized fitting, as only one DM is used to correct the whole FoV. For a DM conjugated to the telescope pupil, ${\bf P_{\beta_j}^{DM}}$ simply becomes the identity matrix, and the projection term then writes as:
\begin{equation}
{\bf P_{opt}} = \langle {\bf P_{\beta_j}^L} \rangle_{\beta}
\end{equation}
it only represents an average over all the direction of optimization. 
An example of the expected performance with GLAO is presented in Fig. \ref{moaoglao} (bottom).
We see that the performance is limited to few \% of SR, but it is almost uniform in the field, and the corrected area is larger than for the MOAO correction. For this example, the error due to generalized fitting at the center of the FoV can be as high as a factor of 5.\\

\subsection{MCAO}
To reduce the projection error, one must use more DMs optically conjugated to the layers in altitude.

\subsubsection{Generalized fitting}
\begin{figure}[h!]
  \begin{center}
   \begin{tabular}{c}
          \includegraphics[height=6.45cm]{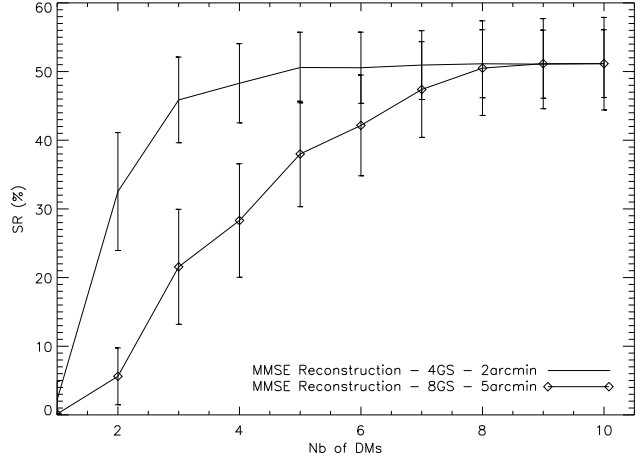} 
          \end{tabular}
   \end{center}
   \caption{Mean SR over the field in function of the number of DMs. Error bars represent the standard deviation of the SR.}
    \label{gfitting}
   \end{figure}

When several DMs are used to perform the correction, the projection term is given by Eq.\ref{popt}. For a given optimized field, the generalized fitting error only depends on the number of DMs \cite{Rigaut-p-00a, Tokovinin-a-00a}. The more DMs used to correct the volume, the better the match between mirrors position and turbulent layers, the better the performance in the field. \\
To illustrate the effect of the generalized fitting error, in Fig. \ref{gfitting} we plot the mean SR over the field (computed over the 169 directions) as a function of the number of DMs. Error bars represent the standard deviation of the SR in the field. Altitudes of the DMs are defined by the altitudes of the equivalent layers, following the same procedure as in Sect. \ref{nboflayer}. Results are shown for the 4GS and 8GS cases using an MMSE reconstruction.\\
Fig. \ref{gfitting} shows that with the 4GS (medium FoV) case, 3 DMs are sufficient to obtain almost an optimal performance. However, for the 8GS (large FoV) case, this number increase to 8 DMs. These results are very similar to those obtained in Fig. \ref{nboflayer}. Indeed, as introduced in Fusco et al. \cite{Fusco-a-01a}, the generalized fitting error is close to a model error on the number of layers. 

\subsubsection{Impact of different reconstructors}
\label{sec_reconsmcao}
\begin{figure}[h!]
  \begin{center}
   \begin{tabular}{c}
          \includegraphics[height=11cm]{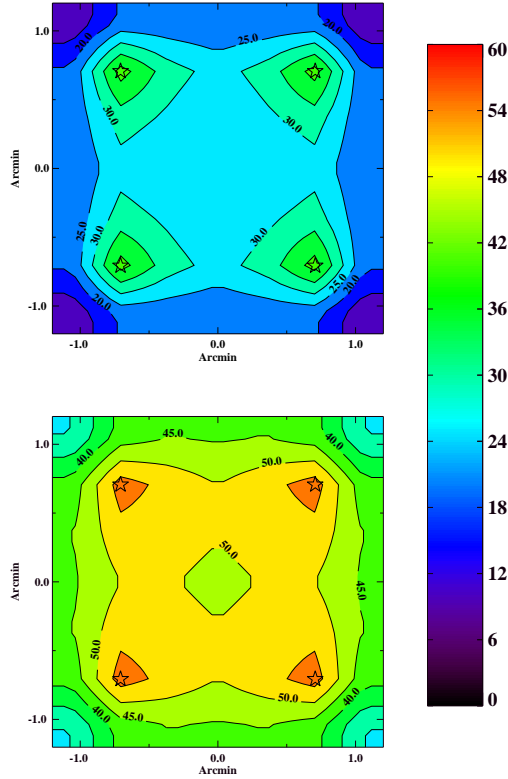} 
          \end{tabular}
   \end{center}
   \caption{SR maps for an MCAO working with 3DMs at [0, 3.5, 9.3]km. Comparison between TLSE (top) and MMSE (bottom) reconstructors. SR[min, max, mean, standard deviation] = [10\%, 42\%, 25\%, 6\%] for the TLSE and [30\%, 57\%, 47\%, 6\%] for MMSE}
    \label{mmsevslse}
   \end{figure}

We now want to investigate the gain of the regularized reconstructor (MMSE) compared to the TLSE one. 
We use the 4GS case with 3DMs located respectively at [0, 3.5, 9.3]km to limit the impact of generalized fitting (see Fig. \ref{gfitting}). 
In Fig. \ref{mmsevslse} we show the SR map for the TLSE (top) and MMSE (bottom) approach.
We find the MMSE provides a performance $\sim$1.2 times better in mean SR: there is a significant gain to use a regularized algorithm compare to a crude TLSE one. Particularly in the corners of the field where the MMSE extrapolate the phase estimation, whereas the TLSE becomes very sensitive to unseen frequencies. Note that in the TLSE case, another choice of threshold could make the performance better on the GS (as good as the MMSE reconstruction), at the price of decreasing the performance in the field.

\subsubsection{Equivalent Layers}
\label{sec_equivalentlays}
\begin{figure}[h!]
  \begin{center}
   \begin{tabular}{c}
          \includegraphics[height=11cm]{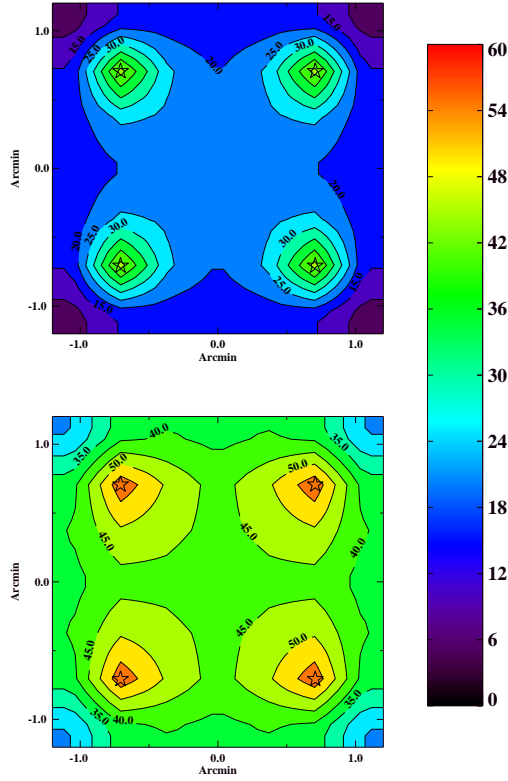} 
          \end{tabular}
   \end{center}
   \caption{SR MMSE in the equivalent layers approach: turbulent volume is reconstructed only at DMs altitudes. {\bf Top} TLSE reconstructor SR[min, max, mean, standard deviation] = [7\%, 47\%, 21\%, 7\%] {\bf Bottom:} MMSE reconstructor. SR[min, max, mean, standard deviation] = [24\%, 58\%, 42\%, 6\%]}
    \label{reallse}
   \end{figure}

Another classical approach in MCAO is to reconstruct the turbulent volume only at the DMs altitudes. No projection onto the DMs is then required and ${\bf P_{opt}} = {\bf Id}$. Following our approach, this is nothing else that a model error on the number of layers as described in Sect. \ref{nboflayer}. 
In Sect. \ref{nboflayer} we focused on the effect of this error at the center of the field, we now want to evaluate the impact of this approach for the performance in the whole field. We choose to simulate a system working with 3DMs as described in Sect.\ref{sec_reconsmcao}. In Fig. \ref{reallse} we compare the respective performance obtained for the TLSE (top) and the MMSE (bottom) reconstructors. 

Similarly to the results of Sect. \ref{sec_reconsmcao}, we first find that the MMSE approach provides a performance $\sim$1.2 times better than the TLSE one. 
Another interesting exercise is to compare the results of Fig.\ref{reallse} with those of Fig.\ref{mmsevslse}.
Doing so, we first find that the mean SR are very close for both approaches: a reconstruction performed directly on the DMs only gives a 5\% less performance in SR. This is indeed consistent with the results obtained in Fig. \ref{nlays} for the center of the field. Note that this errors strongly depends on the turbulence profile, and different turbulent conditions would give different results. In any cases, the main difference between Fig.\ref{reallse} and Fig.\ref{mmsevslse} is that the correction is no more uniform over the field when one uses a reconstruction directly onto the DMs. Indeed, the optimization of the performance for specific directions comes from the projection term ${\bf P_{opt}}$. To enable an optimization of the performance in the field, one must then reconstruct the turbulent volume on more layers than DMs. The other main draw-back of a reconstruction directly on the DMs is that the corresponding model error is very sensitive for large FoV systems (see Sect. \ref{nboflayer}). For instance, the same study for the 8GS configuration only gives a mean SR of 15\% for the MMSE reconstructor. The main advantage of the EL approach is that the tomography is much more simplified compared to the full MCAO presented in Sect. \ref{sec_reconsmcao}.

\section{Conclusion}

We have first presented a matrix formalism for the Fourier modeling of any WFAO systems.
Our Fourier tool includes all the specifies of the WFAO systems such as tomography, number/positions of DMs, model/statistical errors.
 The Fourier approach is interesting because it allows a fast and easy exploration of the broad parameter space, as well as a detailed comprehension of the underlying physical phenomena. For ELTs studies, it offers a fine and accurate modeling tool, able to provide the end-product PSFs.\\
Based on this Fourier tool, we have explored three main issues of any WFAO system, respectively: unseen frequencies, model/statistical errors and projection errors. Our goal was not to derive numerical results, but rather to point out general trends shared by the tomographic systems. 
We first illustrated how the GS geometry set the amount of unseen frequencies. The repartition of unseen frequencies is essential to understand the additional errors: they represent the roots where additional errors are growing.\\
Our exploration of model/statistical errors draw several limitations shared by all WFAO systems.
First, we retrieve that the MMSE approach is always more accurate and robust than the best TLSE one.
Then, that the tomographic reconstruction is robust to statistical errors, but is very sensitive to model errors. 
For medium FoV systems, a good knowledge of any statistical/atmospheric priors is not essential. However, for large FoV systems the sensitivity to turbulent distribution errors becomes significant. This implies that large FoV concepts would require a large number of GS to limit unseen frequencies, or a good monitoring of turbulent conditions.\\
We also derive few rules for a robust control when system and atmospheric conditions are not perfectly known. In such cases, it is advice to (i) overestimate noise variance, (ii) undererestimate turbulence strength, (iii) tends to a constant turbulence profile strength, (iv) use more layers in the reconstruction. Finally, we give a first insight into typical performance of the future WFAO systems for ELTs, and we show that regularized tomographic algorithms are essential to provide a significant gain in performance for these future systems.

\paragraph{Acknowledgments}
This work was supported in part  by: (i) ONERA (the French Aerospace Lab) ; (ii) the French Agence Nationale de la Recherche (ANR) program 06-BLAN-0191; (iii) The European Southern Observatory, phase A study of a Wide Field, multi-IFU near IR Spectrograph and AO system for the E-ELT; (iv) The European Community (Framework Programme 7, E-ELT Preparation, contract No INFRA-2.2.1.28.
Authors are grateful to G\'erad Rousset, Cyril Petit, Fran\c{c}ois Ass\'emat and Laurent Jolissaint for fruitful discussions.

\end{document}